\documentclass[12pt]{article}
\usepackage{amssymb}
\usepackage{amsmath}
\usepackage{latexsym}
\usepackage{epsfig}
\usepackage{dsfont,eucal,mathrsfs}

\newif{\ifcomentarios}
\comentariosfalse

\newtheorem{theorem}{Theorem}

\newtheorem{definition}[theorem]{Definition}
\newtheorem{proposition}[theorem]{Proposition}

\newtheorem{example}[theorem]{Example}

\newtheorem{remark}[theorem]{Remark}

\newcommand{\dint}{\displaystyle\int}

\renewcommand{\mathbf}{\boldsymbol}
\renewcommand{\mathcal}{\mathscr}

\begin{document}

\author{\textbf{Leonardo F. Guidi}\thanks{%
Supported by FAPESP under grant $\#98/10745-1$. E-mail: \textit{\
guidi@if.usp.br}. \ \ \ \ \ \ } \quad \&\quad \textbf{Domingos H. U.
Marchetti} \thanks{%
Partially supported by CNPq and FAPESP. E-mail: \textit{\ marchett@if.usp.br}%
} \\
Instituto de F\'{\i}sica\\
Universidade de S\~{a}o Paulo \\
Caixa Postal 66318\\
05315 S\~{a}o Paulo, SP, Brasil}
\title{Convergence of\ the Mayer Series via Cauchy Majorant Method with
Application to the Yukawa Gas in the Region of Collapse}
\date{}
\maketitle

\begin{abstract}
We construct majorant functions $\Phi (t,z)$ for the Mayer series of
pressure satisfying a nonlinear differential equation of first order which
can be solved by the method of characteristics. The domain $\left\vert
z\right\vert <\left( e\tau \right) ^{-1}$ of convergence of Mayer series is
given by the envelop of characteristic intersections. For non negative
potentials we derive an explicit solution in terms of the Lambert $W$%
--function which is related to the exponential generating function $T$ of
rooted trees as $T(x)=-W(-x)$. For stable potentials the solution is
majorized by a non negative potential solution. There are many choices in
this case and we combine this freedom together with a Lagrange multiplier to
examine the Yukawa gas in the region of collapse. We give, in this paper, a
sufficient condition to establish a conjecture of Benfatto, Gallavotti and
Nicol\'{o}. For any $\beta \in \lbrack 4\pi ,8\pi )$, the Mayer series with
the leading terms of the expansion omitted (how many depending on $\beta $)
is shown to be convergent provided an improved stability condition holds.
Numerical calculations presented indicate this condition is satisfied if few
particles are involved.
\end{abstract}

\section{Introduction}

\setcounter{equation}{0} \setcounter{theorem}{0}

One of main investigations in equilibrium statistical mechanics concerning
the determination of phase diagrams is to define regions of the parameter
space in which a phase transition does not occur. According to Lee--Yang
Theorem \cite{YL,LY} theses regions are free of zeros of the partition
function which, for lattice models with attractive pair potentials, are
located in the unit circle of the complex activity plane. Another way of
determining whether the phase diagram is free of singularities is given by
Mayer series. Although the theory based on these series does not describe
the equation of state beyond the condensate point, it is able, however, to
deal with continuum as well as lattice space for a larger class of
admissible pair potentials.

The domain of convergence of Mayer series for classical gas with stable
two--body summable potential can be established using various different
techniques (see Ruelle \cite{R}, Brydges \cite{B} and references therein).
There are cases, including attractive potential at short distances and
Yukawa gas at inverse temperature $\beta $ up to $4\pi $, which require the
use of finitely \cite{GM} or infinitely \cite{Be,I} many Mayer series
iterations. The present investigation uses the flow equation of Ursell
functions introduced by Brydges and Kennedy \cite{BK,BK1} in order to obtain
previously known and some new results on the convergence of Mayer series
using methods of partial differential equations (PDE).

The radius of convergence of a Mayer series is often determined by a
\textquotedblleft non--Physical\textquotedblright\ singularity in the
complex activity $z$--plane. In the present article we shall write a
majorant series $\Phi (t,z)$ for the pressure $p(\beta ,z)$ of a gas which
satisfies a nonlinear partial differential equation of first order. The
\textquotedblleft non--Physical\textquotedblright\ singularity can be
recognized as the location, in the phase space $\Omega =\left\{ \left(
t,z\right) \in \mathbb{R}^{2}:t\in \left[ t_{0},t_{1}\right] ,~z\geq
0\right\} $, of characteristics intersection ($t$ is eventually related with 
$\beta $). As in Tonks' model of one--dimensional rods with hard core
interaction, the singularity of $\Phi (t,z)$ is manifested by the $W$%
--Lambert function, an indication that this singularity is of combinatorial
nature \cite{K}.

In the application to the Yukawa gas, the variable $t\in \left[ t_{0},0%
\right] $ parameterizes the decomposition of the Yukawa potential into
scales, with $t_{0}$ being related with the short distance cutoff, which is
removed if $t_{0}\rightarrow -\infty $. There is a conjecture that the
radius of convergence of the corresponding Mayer series remains finite as $%
t_{0}\rightarrow -\infty $ for $\beta \in \lbrack 4\pi ,8\pi )$ if one
subtracts the leading terms of the expansion, how many depending on $\beta $%
. Stated as an open problem in \cite{Be} and previously formulated in \cite%
{BGN,GN}, the conjecture will be referred here as BGN's Conjecture.
Brydges--Kennedy \cite{BK} have proved convergence of Mayer series for $4\pi
\leq \beta <16\pi /3$ with the $O\left( z^{2}\right) $ term omitted and
shown how it could be extended up to $\beta <6\pi $ if an improved stability
condition for three particles holds. In the present work we establish BGN's
conjecture up to $\beta <8\pi $, assuming a mild improvement on the
stability condition for the energy of non--neutral $n$--particle
configurations. As our calculation indicates, this condition can be easily
verified numerically if $n$ is not too large since the minimal value
attained by the energy maintains a comfortable distance from the lower bound
needed to establish convergence.

At inverse temperature $8\pi \left( 1+(2r+1)^{-1}\right) ^{-1}$, $%
r=1,2,\ldots $, neutral multipoles of order $\leq 2r$ become unstable and
collapse if the cutoff is removed. Even coefficients of the Mayer series up
to order $2r$ diverges as $t_{0}\rightarrow -\infty $ and they remain finite
otherwise because there occur cancellations when unstable multipoles
interact with another particle. Our proof of convergence has two distinct
parts. In the first, the solution $\Phi ^{(k)}(t,z)$ of a nonlinear first
order PDE modified by Lagrange multipliers, with $\Phi ^{(k)}(t_{0},z)=z$,
is shown to be analytic in a disc whose radius remains positive as $%
t_{0}\rightarrow -\infty $ for any $\beta <\beta _{k}=8\pi \left(
1+k^{-1}\right) ^{-1}$. In the second part, $\Phi ^{(k)}$ is shown to
majorize the Mayer series with even coefficients up to order $k$ subtracted.
The two ingredients here are an improved stability bound for non--neutral
configurations and translational invariance. All estimates are done in the
infinite volume limit to avoid boundary problems. As the Yukawa potential
decays exponentially fast this choice, although not relevant from the point
of view of thermodynamics, is suitable to exhibit the necessary
cancellations right at the beginning.

This paper is organized as follows. Section \ref{GI} contains a review of
Gaussian integrals, including their relation to the partition function of
classical gases, and the statements of our results in Theorems \ref%
{convergence} and \ref{singularity}. Section \ref{UFE} gives an alternative
derivation of Ursell integral equation of Brydges--Kennedy. The integral
equation is the starting point for the Majorant differential equation
presented in Section \ref{MDE}. This first order nonlinear partial
differential equation is solved in Section \ref{MC} by the method of
characteristics for non negative and stable potentials, concluding the proof
of Theorem \ref{convergence}. Section \ref{AA} and \ref{Cjt} give an
application of the previous section to the ultraviolet problem of the
two--dimension Yukawa gas. Section \ref{AA} reestablishes previously known
results via our method. In Sections \ref{Cjt} we prove BGN's Conjecture
under the assumption of an improved stability condition (Theorem \ref%
{singularity}) which has been verified numerically for $n=3,4$ and $5$ (see
Remark \ref{isc}).

\section{Gaussian Integrals and Summary of Results\label{GI}}

\setcounter{equation}{0} \setcounter{theorem}{0}

To fix our notations, we need some facts on Gaussian integrals and their
relation to the grand partition function (see references \cite{FS,Br,BM},
for details).

The state of a particle $\xi =\left( x,\sigma \right) $ consists of its
position $x$ and some internal degrees of freedom $\sigma $. \textit{%
Examples:} $1.$ For one--component lattice gases, $\xi =x\in \mathbb{Z}^{d}$%
; $2.$ For a two--component continuous Coulomb gas, $\xi =\left( x,\pm
e\right) $ where $x\in \mathbb{R}^{d}$ and $e$ is the electrical charge; $3.$
For dipole lattice gas, $\xi =\left( x,\pm e_{j}\right) $\ where $x\in 
\mathbb{Z}^{d}$ and $e_{j}$ is the canonical unit vector.

The particles are constrained to move in a finite subset $X$ of $\mathbb{Z}%
^{d}$ (or a compact region of $\mathbb{R}^{d}$). The lattice is preferred
for pedagogical reason but we shall also consider particles in a continuum
space. $S$ is in general a compact space and we set $\Lambda =X\times S$. By
the triple $\left( \Omega ,\Sigma ,d\rho \right) $ we mean the measure space
of a single particle with $\Omega =\mathbb{R}^{d}\times S$ , $\Sigma $ a
sigma algebra generated by the cylinder sets of $\mathbb{R}^{d}\times S$ and 
$d\rho (\xi )=d\mu (x)\,d\nu (\sigma )$ a product of positive measures ($%
d\mu (x)=\displaystyle\sum\nolimits_{n\in \mathbb{Z}^{d}}\delta
(x-n)\,d^{d}x $ is a counting or Lebesgue $d\mu (x)=\,d^{d}x$ measure times
a probability measure $d\nu $ on $S$).

The grand canonical partition function is given by 
\begin{equation}
\Xi _{\Lambda }(\beta ,z)=\sum_{n=0}^{\infty }\frac{z^{n}}{n!}\int_{\Lambda
^{n}}d^{n}\rho (\mathbf{\xi })\,\psi _{n}(\mathbf{\xi })  \label{partition}
\end{equation}%
where, for each state $\mathbf{\xi }=\left( \xi _{1},\ldots ,\xi _{n}\right) 
$ of $n$--particles with activity $z$, 
\begin{equation}
\psi _{n}(\mathbf{\xi })=\,e^{-\beta U_{n}(\mathbf{\xi })}  \label{boltz}
\end{equation}%
is the associate Boltzmann factor,%
\begin{equation}
U_{n}(\mathbf{\xi })=\sum_{i<j}\vartheta (\xi _{i},\xi _{j})  \label{Un}
\end{equation}%
the interaction energy and $\beta $ is the inverse temperature. The
two--body potential $\vartheta :\Lambda \times \Lambda \longrightarrow 
\mathbb{R}$ is a jointly measurable function and we use, sometimes, the
abbreviation $\vartheta _{ij}:=\vartheta (\xi _{i},\xi _{j})$. The finite
volume pressure $p_{\Lambda }(\beta ,z)$ of a classical gas is defined by 
\begin{equation}
p_{\Lambda }(\beta ,z)=\frac{1}{\beta \left\vert \Lambda \right\vert }\ln
\Xi _{\Lambda }(\beta ,z)\,.  \label{pressure}
\end{equation}

Without loss, $\vartheta $ is defined in the whole space $\mathbb{Z}^{d}$
(or $\mathbb{R}^{d}$) - the Green's function of some differential equation
with free (insulating) boundary conditions - and it is assumed to be of the
form 
\begin{equation}
\beta \vartheta (\xi ,\xi ^{\prime })=\,\sigma \sigma ^{\prime }v\left(
x-x^{\prime }\right)   \label{w}
\end{equation}%
for some real--valued measurable functions $v$ satisfying:

\begin{enumerate}
\item \textit{Regularity at origin} $\left\vert v(0)\right\vert \leq
2B<\infty $;

\item \textit{Summability} 
\begin{equation}
\left\Vert v\right\Vert :=\int \,d\mu (x)\,\left\vert v(x)\right\vert
<\infty ~;  \label{norm}
\end{equation}

\item \textit{Positivity} 
\begin{equation}
\sum_{1\leq i,j\leq n}\overline{z}_{i}\,z_{j}\,v\left( x_{i}-x_{j}\right)
\geq 0\,,  \label{pos}
\end{equation}%
for any numbers $x_{1},\ldots ,x_{n}\in \mathbb{R}^{d}$ and $z_{1},\ldots
,z_{n}\in \mathbb{C}$, $n=1,2,\ldots $.
\end{enumerate}

Note these assumptions ensure that $\vartheta $ is a stable potential, i.e.,
choosing $z_{1}=\cdots =z_{n}=1$ in (\ref{pos}), 
\begin{equation}
\beta U_{n}(\mathbf{\xi })\geq -\frac{1}{2}\sum_{1\leq i\leq n}\sigma
_{i}^{2}\left\vert v(0)\right\vert \geq -Bn  \label{B}
\end{equation}%
holds with $B\geq 0$ independent of $n$ (see Section $3.2$ of Ruelle \cite{R}%
, for further comments).

Since $v$ is a positive definite symmetric function, it defines a Gaussian
measure $\mu _{v}$ with mean $\displaystyle\int d\mu _{v}(\phi )\phi (x)\,=0$
and covariance $\int d\mu _{v}(\phi )\,\phi (x)\,\phi (x^{\prime
})=v(x-x^{\prime })$. Introducing the inner product 
\begin{equation*}
\left( \phi ,\eta \right) =\int d\mu (x)\,\phi (x)\,\eta (x)
\end{equation*}%
and the indicator function 
\begin{equation}
\eta _{n}(\cdot )=\eta (\mathbf{\xi };\cdot )=\sum_{j=1}^{n}\sigma
_{j}~\delta \left( \cdot -x_{j}\right)  \label{ind}
\end{equation}%
of a $n$--state $\mathbf{\xi }=\left( \xi _{1},\ldots ,\xi _{n}\right) $, we
have 
\begin{equation}
\psi _{n}(\mathbf{\xi })=\int d\mu _{v}\left( \phi \right) \,:e^{i\left(
\phi ,\eta _{n}\right) }:_{v}  \label{boltz1}
\end{equation}%
where $:\,:_{v}$ is the normal ordering with respect to the covariance $v$: 
\begin{equation}
:e^{i\sigma \phi (x)}:_{v}=\exp \left\{ \frac{\sigma ^{2}}{2}v\left(
0\right) \right\} \,e^{i\sigma \phi (x)}\,.  \label{order}
\end{equation}

Substituting (\ref{boltz1}) into (\ref{partition}) yields 
\begin{equation}
\Xi _{\Lambda }(\beta ,z)=\int d\mu _{v}\left( \phi \right) \,\exp \left\{
V_{0}(\phi )\right\}  \label{part}
\end{equation}%
where 
\begin{equation}
V_{0}(\phi )=z\int_{\Lambda }d\rho (\xi )\,:e^{i\sigma \phi (x)}:_{v}.
\label{V0}
\end{equation}

\begin{example}
\label{example}

\begin{enumerate}
\item $V_{0}(\phi )=z\displaystyle\sum_{x\in X}:e^{i\phi (x)}:_{v}$ for
one--component lattice gas;

\item $V_{0}(\phi )=z\displaystyle\int_{X}d^{d}x:\cos \phi (x):_{v}\,$\ for
standard Coulomb or Yukawa gas;

\item \label{ex3}$V_{0}(\phi )=z\displaystyle\sum_{x\in X}\displaystyle%
\int_{\left\vert \sigma \right\vert =1}d^{d}\sigma :\cos \left( \sigma \cdot
\nabla \phi (x)\right) :_{v}$ for dipole lattice gas.
\end{enumerate}
\end{example}

In Example $2$ we have respectively, $v=(-\Delta )^{-1}$ and $v=\left(
-\Delta +1\right) ^{-1}$, the fundamental solution of Laplace and Yukawa
equations. The potential of two dipoles, of moment $\sigma $ and $\sigma
^{\prime }$, in Example $3.$ is given by $v=(\sigma \cdot \nabla )(\sigma
^{\prime }\cdot \nabla ^{\prime })(-\Delta )^{-1}$, second difference of the
fundamental solution of discrete Laplace equation. $V_{0}$ in Examples $2$
and $3$ contain cosine function instead exponential. This form expresses the
pseudo--charge symmetry required for the existence of thermodynamic limit $%
\lim_{\Lambda \nearrow \mathbb{R}^{d}(\mathbb{Z}^{d})}p_{\Lambda }=p$ of
respective systems (see, e.g., \cite{S}).

The potentials in Examples $2$ and $3$ do not fulfill all requirements:
Coulomb and dipole potentials do not satisfy summability whereas the Yukawa
potential fails to be regular at origin. To circumvent these problems, a one
parameter smooth decomposition of covariance $v(t,x)$ may be introduced so
that $v(t_{0},x)=0$ and $\displaystyle\int_{t_{0}}^{t_{1}}\left\Vert \dot{v}%
(s,\cdot )\right\Vert ds<\infty $ holds for any finite interval $\left[
t_{0},t_{1}\right] $. This allows us to approach the ultraviolet problem of
Yukawa gas by letting $t_{0}\rightarrow -\infty $ maintaining $t_{1}$ fixed
(e.g. $t_{1}=0$) in Example $2$, and the infrared problem of dipole gas in
Example $3$ as the limit $t_{1}\rightarrow \infty $ with $t_{0}=0$ fixed.
The application in the present work deals only with the former problem.

In all examples $V_{0}(\phi )$ is a bounded perturbation\ to the Gaussian
measure $\mu _{v}$ satisfying $\left\vert V_{0}(\phi )\right\vert \leq
z\left\vert \Lambda \right\vert $. Brydges and Kennedy \cite{BK} have used
the\ so called sine--Gordon representation (\ref{part}) to derive a PDE in
the context of Mayer expansion. Although based in their work, our PDE
approach in the present article is remarkably different.

It is convenient to write a flow 
\begin{equation}
V_{\Lambda }\left( t,z;\phi \right) =\ln \left( \mu _{v(t)}\ast \exp \left\{
V_{0}(\phi )\right\} \right)  \label{V}
\end{equation}%
starting from $V_{\Lambda }\left( 0,z;\phi \right) =V_{0}(\phi )$ where $\mu
_{v}\ast F$ denotes convolution of $F$ by the Gaussian measure $\mu _{v}$
which, by the Wick theorem, is given for the one--component lattice case by 
\begin{equation}
\int d\mu _{v}(\zeta
)\,F(\phi +\zeta )
= \exp \left\{ \frac{1}{2}\sum_{x,y\in \mathbb{Z}^{d}}v\left( t,x-y\right) 
\frac{\partial ^{2}}{\partial \phi (x)\partial \phi (y)}\right\} F(\phi
)\,. \label{wick}
\end{equation}
To derive the standard Mayer expansion, the covariance $v(t)$ is chosen to
be any $\mathcal{C}^{1}$ parameterization such that $v(t_{0})=0$ and $%
v(t_{1})=v$. The parameterization involving multiscales, $\dot{v}(t)$ is
required to be of the form (\ref{w}) satisfying properties $1-3$ with $\dot{B%
}=\dot{B}(t)<\infty $ for all $t\in \left[ t_{0},t_{1}\right] $. $v(t)$ is
said to be admissible if fulfills all these properties. Note that $%
V_{0}(\phi )$ also depends on $t$ through the normal ordering $:\,:_{v(t)}$.

Equation (\ref{V}) can be written as 
\begin{equation}
e^{V_{\Lambda }\left( \beta ,z,\phi \right) }=\Xi _{\Lambda }(\beta ,ze^{i%
\mathbf{\phi }})  \label{V1}
\end{equation}%
where $\Xi _{\Lambda }(\beta ,\mathbf{w})$ generalizes the grand partition
function by including a complex valued activity $w=w(\xi )$ which depends on
the particle state: 
\begin{equation}
\Xi _{\Lambda }(\beta ,\mathbf{w})=\sum_{n=0}^{\infty }\frac{1}{n!}%
\int_{\Lambda ^{n}}d^{n}\rho (\mathbf{\xi })\,w_{n}(\mathbf{\xi })\,\psi
_{n}(\mathbf{\xi })  \label{part1}
\end{equation}%
with $w_{n}(\mathbf{\xi })=\displaystyle\prod_{j=1}^{n}w(\xi _{j})$ and (\ref%
{partition}) is recovered if $w(\xi )=z$.

We also have 
\begin{equation}
\beta \left\vert p_{\Lambda }(\beta ,z)\right\vert =\frac{1}{\left\vert
\Lambda \right\vert }\left\vert V_{\Lambda }\left( t_{1},z,0\right)
\right\vert \leq \sup_{\phi }\frac{1}{\left\vert \Lambda \right\vert }%
\left\vert V_{\Lambda }\left( t_{1},z,\phi \right) \right\vert ~  \label{ppp}
\end{equation}%
and, as $t$ goes to $t_{0}$ (the infinite temperature limit), 
\begin{equation*}
\lim_{t\rightarrow t_{0}}V_{\Lambda }\left( t,z,0\right)
=V_{0}(0)=z\left\vert \Lambda \right\vert \,~.
\end{equation*}%
The flow $\left\{ V_{\Lambda }(t,z,\phi )/\left\vert \Lambda \right\vert
:0\leq t\leq 1\right\} $ provides us an upper bound for the pressure of
gases described above, starting from the pressure of an ideal gas.

The results presented in this article can be summarized as follows.

\begin{theorem}
\label{convergence}Let $\Phi =\Phi (t,z)$ be the classical solution of 
\begin{equation}
\Phi _{t}=\frac{1}{2}\Gamma \left( z\Phi _{z}\right) ^{2}+\dot{B}\left(
z\Phi _{z}-z\right)  \label{eq}
\end{equation}%
defined in $t_{0}<t\leq t_{1}$, $z\geq 0$ with $\Phi (t_{0},z)=z$ where $%
\Gamma =\Gamma (t)=\left\Vert \dot{v}(t)\right\Vert $, $B=B(t)$ is defined
by (\ref{B}) with $v$ replaced by $v(t)$. Then%
\begin{equation*}
\beta p_{\Lambda }(\beta ,z)\leq \Phi (t_{1},z)
\end{equation*}%
holds uniformly in $\Lambda $. The solution $\Phi $ has a power series 
\begin{equation*}
\Phi (t,z)=z+\sum_{n=2}^{\infty }A_{n}(t)~z^{n}
\end{equation*}%
with $A_{n}(t)\geq 0$, convergent in the open disc $D=\left\{ \left\vert
z\right\vert <\left( e\tau \right) ^{-1}\right\} $ where 
\begin{equation*}
\tau (t_{0},t)=\inf \left( \int_{t_{0}}^{t}\Gamma (s)\exp \left\{
2\int_{s}^{t}\dot{B}(s^{\prime })~ds^{\prime }\right\} ds,~\exp \left\{
\int_{t_{0}}^{t}\dot{B}(s^{\prime })~ds^{\prime }\right\}
\int_{t_{0}}^{t}\Gamma (s)ds\right)
\end{equation*}%
and is majorized by 
\begin{equation}
\Phi (t,z)\leq \frac{-1}{\tau }\left( W\left( -z\tau \right) +\frac{1}{2}%
W^{2}\left( -z\tau \right) \right)  \label{Phitz}
\end{equation}%
where $W(x)$ is the Lambert $W$--function defined as the inverse of $%
f(W)=We^{W}$.
\end{theorem}

\begin{theorem}
\label{singularity}Given an odd number $k\in \mathbb{N}$, suppose an
improved stability condition 
\begin{equation}
\dot{U}_{n}(t;\mathbf{\xi })\geq -(n-\delta _{n})\dot{B}(t)~  \label{stb1}
\end{equation}%
holds for all $n$--particle states $\mathbf{\xi }=(\mathbf{\sigma },\mathbf{x%
})$ such that $\displaystyle\sum_{i=1}^{n}\sigma _{i}\neq 0$ with%
\begin{equation*}
\delta _{n}>\frac{1}{n}~
\end{equation*}%
and $n\leq k$. Then, the pressure $p\left( \beta ,z\right) $ of Yukawa gas,
with even Mayer coefficients up to order $k$ subtracted, exists and has a
convergent power series provided 
\begin{equation*}
\beta <\beta _{k}\qquad \mathrm{and}\qquad \left\vert z\right\vert <\frac{%
\beta _{k}-\beta }{\beta _{k}\beta e}~
\end{equation*}%
where $\beta _{k}=8\pi \left/ \left( 1+k^{-1}\right) \right. $ is the $k$%
--th threshold.
\end{theorem}

Theorems \ref{convergence} and \ref{singularity} will be proven in Sections %
\ref{MC} and \ref{Cjt}, respectively, by majorizing the solution of (\ref{eq}%
), possibly modified by a Lagrange multiplier $L$, by an explicitly solution
of (\ref{eq}) with $B=0$ and $\Gamma $ replaced by $\Gamma /f$ for a
conveniently chosen function $f$ .

\section{Ursell Flow Equation\label{UFE}}

\setcounter{equation}{0} \setcounter{theorem}{0}

The power series of the pressure (\ref{pressure}) in the activity $z$ is
called Mayer series: 
\begin{equation}
\beta p_{\Lambda }(\beta ,z)=\sum_{n=1}^{\infty }b_{\Lambda ,n}\,z^{n}\,~.
\label{p}
\end{equation}%
The Mayer coefficient $b_{\Lambda ,n}$ has an expression analogous to (\ref%
{partition}), 
\begin{equation}
\left\vert \Lambda \right\vert \,b_{\Lambda ,n}=\frac{1}{n!}\int_{\Lambda
^{n}}d^{n}\rho (\mathbf{\xi })\,\psi _{n}^{c}(\mathbf{\xi })\,  \label{b}
\end{equation}%
where, for each state $\mathbf{\xi }=\left( \xi _{1},\ldots ,\xi _{n}\right)
\in \left( \mathbb{R}\times S\right) ^{n}$ of $n$--particles, $\psi _{n}^{c}(%
\mathbf{\xi })$ is the Ursell functions of order $n$. In terms of Mayer
graphs, 
\begin{equation}
\psi _{n}^{c}(\mathbf{\xi })=\sum_{G}\prod_{\left( i,j\right) \in G}\left(
e^{-\beta \vartheta _{ij}}-1\right)  \label{ursell}
\end{equation}%
where $G$ runs over all connected graphs on $I_{n}=\left\{ 1,\ldots
,n\right\} $ ($G$ is connected if for any two vertex $k,\,l\in I_{n}$ there
is a path of edges $\left( i,j\right) \in I_{n}\times I_{n}$ in $G$
connecting $k$ to $l$).

Equation (\ref{ursell}) is not useful to investigate whether (\ref{p}) is a
convergent series since $\left\vert \sum_{G}1\right\vert >c\left( n!\right)
^{1+\delta }$ holds for some $c,\delta >0$\footnote{%
The number of Mayer graphs with $n$ vertices is $2^{n(n-1)/2}$ and the
number of connected Mayer graphs is of the same order. For this, note that a
tree, the minimal connected graph, has exactly $n-1$ bonds, remaining $%
\dfrac{n^{2}}{2}-\dfrac{3n}{2}+1$ bonds to be freely chosen to be connected.}%
. We shall instead derive a flow equation for $\psi _{n}^{c}(\mathbf{\xi })$
that incorporate the cancellations required to reduce the cardinality of the
sum in (\ref{ursell}) to $O(n!)$.

Given $\mathbf{\xi }=\left( \xi _{j}\right) _{j\geq 1}\in \left( \mathbb{R}%
\times S\right) ^{\infty }$, let $P_{n}\mathbf{\xi }=\left( \xi _{1},\ldots
,\xi _{n}\right) $ denote the projection of $\mathbf{\xi }$ into the
subspace $\left( \mathbb{R}\times S\right) ^{n}$ generated by its $n$ first
components. If $I\subset \mathbb{N}$ is a finite set $I=\left\{ i_{1},\ldots
,i_{n}\right\} $, we set $P_{I}\mathbf{\xi }=\left( \xi _{i_{1}},\ldots ,\xi
_{i_{n}}\right) $. To each sequence $\mathbf{\xi }$ we associate a complex
Boltzmann function 
\begin{equation*}
u(t,\mathbf{\xi })=\exp \left\{ -\sum_{i<j}\sigma _{i}\sigma
_{j}v(t,x_{i}-x_{j})+i\sum_{j}\sigma _{j}\phi (x_{j})\right\}
\end{equation*}%
and define 
\begin{equation}
u_{n}(t,\mathbf{\xi })=u(t,P_{n}\mathbf{\xi })=e^{-\beta U_{n}(\mathbf{\xi }%
)+i(\phi ,\eta _{n})}\,,  \label{un}
\end{equation}%
for $n\geq 2$ where $U_{n}(\mathbf{\xi })=U(P_{n}\mathbf{\xi })$ is the
interaction energy (\ref{Un}) and $\eta _{n}(\cdot )=\eta (P_{n}\mathbf{\xi }%
,\cdot )$ is the indicator function (\ref{ind}). Note that $u_{n}:\mathbb{R}%
_{+}\times \left( \mathbb{R}\times S\right) ^{n}\longrightarrow \mathbb{C}$
is symmetric with respect to the permutation $\pi $ of the index set $%
\left\{ 1,\ldots ,n\right\} $: 
\begin{equation*}
u(t,P_{n}\mathbf{\xi })=u(t,M_{\pi }P_{n}\mathbf{\xi })
\end{equation*}%
where $M_{\pi }:\left( \xi _{1},\ldots ,\xi _{n}\right) \longrightarrow
\left( \xi _{\pi _{1}},\ldots ,\xi _{\pi _{n}}\right) $ is the permutation
matrix corresponding to $\pi $. Note also that $u_{n}(t,\mathbf{\xi })$ can
be written as 
\begin{equation}
u_{n}(t,\mathbf{\xi })=\psi _{n}(\mathbf{\xi })\,e^{i\left( \phi ,\eta
_{n}\right) }=\mu _{v(t)}\ast :e^{i\left( \phi ,\eta _{n}\right) }:_{v(t)},
\label{star}
\end{equation}%
where $\psi _{n}$ is the Boltzmann function (\ref{boltz}), with $v=v(t)$
being an admissible covariance.

Now, let $\mathbf{u}(t)=\left( u_{n}(t,\mathbf{\xi })\right) _{n\geq 0}$
denote the sequence of these functions with $u_{0}(t)=1$ and $u_{1}(t,%
\mathbf{\xi })=e^{i\sigma _{1}\phi (x_{1})}$. At $t=t_{0}$, $v(t_{0})=0$
(absence of interactions) and $\mathbf{u}(t_{0})=\mathbf{u}^{(0)}$ is the
sequence with $u_{n}^{(0)}(\mathbf{\xi })=e^{i\left( \phi ,\eta _{n}\right)
} $ for $n\geq 1$ and $u_{0}^{(0)}(\mathbf{\xi })=1$. Using (\ref{star}), (%
\ref{order}) and (\ref{wick}), a simple computation yields

\begin{proposition}
The sequence $\mathbf{u}(t)=\left( u_{n}(t,\mathbf{\xi })\right) _{n\geq 0}$
is the formal solution of the heat equation 
\begin{equation}
\mathbf{u}_{t}=\frac{1}{2}\Delta _{\dot{v}(t)}\mathbf{u}  \label{heat}
\end{equation}%
with initial condition $\mathbf{u}(t_{0})=\mathbf{u}^{(0)}$, where the
action of $\Delta _{D}$ on $\mathbf{u}$ is a sequence $\Delta _{D}\mathbf{u}$
with 
\begin{equation}
\left( \Delta _{D}u\right) _{n}=\sum_{1\leq i,j\leq n,i\neq
j}D(x_{i},x_{j})\,\frac{\partial ^{2}u_{n}}{\partial \phi (x_{i})\partial
\phi (x_{j})}\,.  \label{delta}
\end{equation}
\end{proposition}

\begin{remark}
\label{deriv}

\begin{enumerate}
\item The derivatives in (\ref{delta}) are partial derivatives with respect
to $\phi $ at the value $x$, even for continuous space. This has to be
contrasted with the functional derivative in (\ref{wick}) whose
correspondence to partial derivative can only be made for lattice space.

\item To emphasize the order of operation, we write $\Delta _{D}=\nabla
\cdot D\nabla $ for the differential operator (\ref{delta}) if $D$ is a
off--diagonal symmetric matrix.
\end{enumerate}
\end{remark}

Let $\mathfrak{S}$ denote the vector space over $\mathbb{C}$ of sequences $%
\mathbf{\psi }=\left( \psi _{n}(\mathbf{\xi })\right) _{n\geq 0}$ of bounded
Lebesgue measurable functions $\psi _{n}(\mathbf{\xi })=\psi (P_{n}\mathbf{\
\xi })$ on $\left( \mathbb{R\times S}\right) ^{n}$ which are symmetric with
respect to the permutation $\pi $ of the index set $\left\{ 1,\ldots
,n\right\} $, with $\psi _{0}(\mathbf{\xi })=\psi \left( P_{\emptyset } 
\mathbf{\xi }\right) \in \mathbb{C}$. We define a product operation $\circ : 
\mathfrak{S}\times \mathfrak{S}\longrightarrow \mathfrak{S}$ \ for any two
sequences of symmetric functions $\mathbf{\psi }_{1},\,\mathbf{\psi }_{2}$
as a sequence 
\begin{equation}
\mathbf{\psi }_{1}\circ \mathbf{\psi }_{2}=\mathbf{\psi }=\left( \psi _{n}( 
\mathbf{\xi })\right) _{n}  \label{prod}
\end{equation}
whose components are given by 
\begin{equation*}
\psi _{n}(\mathbf{\xi })=\sum_{I\subseteq \left\{ 1,\ldots ,n\right\} }\psi
_{1}(P_{I}\mathbf{\xi })\,\psi _{2}(P_{I^{c}}\mathbf{\xi })
\end{equation*}%
where $I^{c}=\left\{ 1,\ldots ,n\right\} \backslash I$. Note that this
operation is well defined and preserves the symmetry. $\mathfrak{S}$ endowed
with the operation $\circ $ form a commutative algebra with the identity $%
\mathbf{1}=\left( \delta _{n}(\mathbf{\xi })\right) _{n\geq 0}$ given by 
\begin{equation*}
\delta (P_{I}\mathbf{\xi })=\left\{ 
\begin{array}{lll}
1 & \mathrm{if} & I=\emptyset \\ 
0 & \mathrm{if} & I\neq \emptyset%
\end{array}
\right. \,.
\end{equation*}

Let $\mathfrak{S}_{0}$ and $\mathfrak{S}_{1}$ denote the subspace and the
affine space of sequences $\mathbf{\psi }=\left( \psi _{n}(\mathbf{\xi }
)\right) _{n\geq 0}$ such that $\psi _{0}(\mathbf{\xi })=0$ and $\psi _{0}( 
\mathbf{\xi })=1$, respectively. We define the algebraic exponential of
sequences $\mathbf{\phi =}\left( \phi _{n}(\mathbf{\xi })\right) _{n\geq 0}$ 
\begin{equation*}
\mathcal{E}xp:\mathfrak{S}_{0}\longrightarrow \mathfrak{S}_{1}
\end{equation*}%
as the following formal series 
\begin{equation*}
\mathcal{E}xp\left( \mathbf{\phi }\right) =\mathbf{1}+\mathbf{\phi }+\frac{1 
}{2}\mathbf{\phi }\circ \mathbf{\phi }+\cdots +\frac{1}{n!}\underset{n- 
\mathrm{times}}{\underbrace{\mathbf{\phi }\circ \cdots \circ \mathbf{\phi }}%
+ }\cdots
\end{equation*}
and the algebraic logarithmic of sequences $\mathbf{\psi }=\left( \psi _{n}( 
\mathbf{\xi })\right) _{n\geq 0}$ 
\begin{equation*}
\mathcal{L}n:\mathfrak{S}_{1}\longrightarrow \mathfrak{S}_{0}
\end{equation*}
as ($\mathbf{\psi }\in \mathfrak{S}_{0}$) 
\begin{equation*}
\mathcal{L}n\left( \mathbf{1}+\mathbf{\psi }\right) =\mathbf{\psi }-\frac{1}{
2}\mathbf{\psi }\circ \mathbf{\psi }+\cdots +\frac{\left( -1\right) ^{n-1}}{%
n }\underset{n-\mathrm{times}}{\underbrace{\mathbf{\psi }\circ \cdots \circ 
\mathbf{\psi }}+}\cdots \,.
\end{equation*}
Note that the following 
\begin{equation*}
\mathcal{L}n\left( \mathcal{E}xp\left( \mathbf{\phi }\right) \right) = 
\mathbf{\phi }\qquad \mathrm{and}\qquad \mathcal{E}xp\left( \mathcal{L}
n\left( \mathbf{1}+\mathbf{\psi }\right) \right) =\mathbf{1}+\mathbf{\psi \,}
\end{equation*}%
holds for any $\mathbf{\phi },$ $\mathbf{\psi }\in \mathfrak{S}_{0}$, so $%
\mathcal{E}xp$ and $\mathcal{L}n$ are algebraic inverse function of each
other. We thus have (for details, see Ruelle \cite{R})

\begin{proposition}
The sequence $\mathbf{\psi }=\left( \psi _{n}\right) _{n\geq 0}$ of
Boltzmann functions given by (\ref{boltz}) $n\geq 2$ with $\psi _{0}=\psi
_{1}(\mathbf{\xi })=1$, is an algebraic exponential of the sequence $\mathbf{%
\psi }^{c}=\left( \psi _{n}^{c}\right) _{n\geq 0}$ of Ursell functions given
by (\ref{ursell}) with $\psi _{0}^{c}=0$ and $\psi _{1}^{c}(\mathbf{\xi })=1$%
: 
\begin{equation}
\mathbf{\psi }=\mathcal{E}xp\left( \mathbf{\psi }^{c}\right) \,.  \label{exp}
\end{equation}
\end{proposition}

\noindent \textit{Proof.} Let $\langle \cdot ,\cdot \rangle :\mathfrak{S}%
\times \mathfrak{S}\longrightarrow \mathbb{C}$ be the inner product 
\begin{equation}
\langle \mathbf{\phi },\mathbf{\psi }\rangle =\sum_{n=0}^{\infty }\frac{z^{n}%
}{n!}\int d^{n}\rho (\mathbf{\xi })\,\phi _{n}(\mathbf{\xi })\,\psi _{n}(%
\mathbf{\xi })\,  \label{inner}
\end{equation}%
and let $\mathbf{\chi }=\left( \chi _{n}(\mathbf{\xi })\right) _{n\geq 0}$
be the sequence of characteristic function of the set $\Lambda ^{n}$: 
\begin{equation*}
\chi _{n}(\mathbf{\xi })=\prod_{j=1}^{n}\chi _{\Lambda }(\xi _{j})
\end{equation*}%
with 
\begin{equation*}
\chi _{\Lambda }(\xi )=\left\{ 
\begin{array}{lll}
1 &  & \mathrm{if}\;\xi \in \Lambda \\ 
0 &  & \mathrm{otherwise}%
\end{array}%
\right. \,.
\end{equation*}%
Using 
\begin{equation*}
\langle \mathbf{\chi },\mathbf{\phi }\circ \mathbf{\psi }\rangle =\langle 
\mathbf{\chi },\mathbf{\phi }\rangle \,\langle \mathbf{\chi },\mathbf{\psi }%
\rangle
\end{equation*}%
($\mathbf{\phi }\longrightarrow \langle \mathbf{\chi },\mathbf{\phi }\rangle
=\langle \mathbf{\chi },\mathbf{\phi }\rangle (z)$ defines an homeomorphism
of $\mathfrak{S}$ into the algebra of formal power series) together with (%
\ref{exp}), (\ref{inner}) and (\ref{partition}), we have 
\begin{equation*}
\Xi _{\Lambda }(\beta ,z)=\langle \mathbf{\chi },\mathbf{\psi }\rangle
(z)=\langle \mathbf{\chi },\mathcal{E}xp\left( \mathbf{\psi }^{c}\right)
\rangle (z)=\exp \left\{ \langle \mathbf{\chi },\mathbf{\psi }^{c}\rangle
(z)\right\}
\end{equation*}%
which, in view of (\ref{p})--(\ref{ursell}), proves the proposition.

\hfill $\Box $

Replacing $\mathbf{\psi }$ in (\ref{exp}) by the complex Boltzmann sequence $%
\mathbf{u}(t)=\left( u_{n}(t,\mathbf{\xi })\right) _{n\geq 0}$, leads to the
complex Ursell sequence $\mathbf{y}(t)=\left( y_{n}(t,\mathbf{\xi })\right)
_{n\geq 0}$ given by 
\begin{equation}
\mathbf{y}(t)=\mathcal{L}n\left( \mathbf{u}(t)\right) ~.  \label{vlnu}
\end{equation}%
Note that, in view of (\ref{vlnu}), (\ref{star}), (\ref{exp}) and the fact 
\begin{equation*}
\left( u\circ u\right) _{n}(t,\mathbf{\xi })=\sum_{I\subseteq \left\{
1,\ldots ,n\right\} }\psi (t,P_{I}\mathbf{\xi })\,\psi (t,P_{I^{c}}\mathbf{\
\xi })e^{i\left( \phi ,\eta _{n}\right) }=\left( \psi \circ \psi \right)
_{n}(t,\mathbf{\xi })e^{i\left( \phi ,\eta _{n}\right) }~,
\end{equation*}%
we have $y_{n}(t,\mathbf{\xi })=\psi _{n}^{c}(t,\mathbf{\xi })\,e^{i\left(
\phi ,\eta _{n}\right) }$. We shall use (\ref{vlnu}) together with (\ref%
{heat}) to derive a partial differential equation for the complex Ursell
functions $y_{n}(t,\mathbf{\xi })$. A differential equation for the Ursell
function $\psi _{n}^{c}(t,\mathbf{\xi })$, given by (\ref{ursell}) with $%
\beta \vartheta \left( \xi _{i},\xi _{j}\right) $ replaced by $\sigma
_{i}\sigma _{j}v(t,x_{i}-x_{j})$, is obtained setting $\phi =0$ at the end
of computations.

The derivative of (\ref{vlnu}) together with Remark \ref{deriv}$.2$, yields 
\begin{eqnarray}
\mathbf{y}_{t} &=&\frac{1}{\mathbf{u}}\circ \mathbf{u}_{t}  \notag \\
&=&\frac{1}{2}\mathcal{E}xp\left( -\mathbf{y}\right) \circ \left( \nabla
\cdot \dot{v}\nabla \mathcal{E}xp\left( \mathbf{y}\right) \right)   \notag \\
&=&\frac{1}{2}\Delta _{\dot{v}}\mathbf{y}+\frac{1}{2}\nabla \mathbf{y}%
\bullet \dot{v}\nabla \mathbf{y}  \label{v}
\end{eqnarray}%
where 
\begin{eqnarray*}
\left( \nabla \mathbf{y}\bullet D\nabla \mathbf{y}\right) _{n}
&=&\sum_{1\leq i,j\leq n,i\neq j}D(x_{i},x_{j})\left( \frac{\partial \mathbf{%
y}}{\partial \phi (x_{i})}\circ \frac{\partial \mathbf{y}}{\partial \phi
(x_{j})}\right) _{n} \\
&=&\sum_{I\subset \left\{ 1,\ldots ,n\right\} }\,\nabla y\left( P_{I}\mathbf{%
\ \xi }\right) \cdot D\nabla y\left( P_{I^{c}}\mathbf{\xi }\right) ~,
\end{eqnarray*}%
from which equation ($6$) of \cite{BK1} is obtained: 
\begin{equation}
\frac{\partial }{\partial t}\psi _{n}^{c}(\mathbf{\xi })=-\dot{U}_{n}(%
\mathbf{\xi })\,\psi _{n}^{c}(\mathbf{\xi })-\frac{1}{2}\sum_{I\subseteq
\left\{ 1,\ldots ,n\right\} }\dot{U}\left( P_{I}\mathbf{\xi };P_{I^{c}}%
\mathbf{\xi }\right) \,\psi ^{c}(P_{I}\mathbf{\xi })\,\psi ^{c}(P_{I^{c}}%
\mathbf{\xi })  \label{diff}
\end{equation}%
satisfying initial condition $\mathbf{\psi }_{0}^{c}=\left( \psi _{0,n}^{c}(%
\mathbf{\xi })\right) _{n\geq 0}$ at $t_{0}$ with 
\begin{equation}
\psi _{0}^{c}\left( P_{I}\mathbf{\xi }\right) =\left\{ 
\begin{array}{lll}
1 &  & \mathrm{if}\;\left\vert I\right\vert =1 \\ 
0 &  & \mathrm{otherwise}%
\end{array}%
\right. \,.  \label{ic}
\end{equation}%
For brevity, we have omitted the time dependence in (\ref{diff}). Here, $%
U_{n}(t,\mathbf{\xi })=U(t,P_{n}\mathbf{\xi })$ is the energy of state $P_{n}%
\mathbf{\xi }$ and 
\begin{equation}
U\left( t,P_{I}\mathbf{\xi };P_{J}\mathbf{\xi }\right) =\sum_{i\in I,\,j\in
J}\sigma _{i}\sigma _{j}v\left( t,x_{i}-x_{j}\right) \,~.  \label{S}
\end{equation}%
The minus sign in the r.h.s. of (\ref{diff}) comes from the derivatives of $%
e^{i\phi (\xi )}$ at $\phi =0$.

Using the variation of constants formula, (\ref{diff}) is equivalent to the
following integral equation 
\begin{equation}
\psi _{n}^{c}(t,\mathbf{\xi })=-\frac{1}{2}\int_{t_{0}}^{t}ds\,e^{-%
\displaystyle\int_{s}^{t}d\tau \,\dot{U}_{n}(\tau ,\mathbf{\xi }%
)}\sum_{I\subseteq \left\{ 1,\ldots ,n\right\} }\dot{U}\left( s,P_{I}\mathbf{%
\xi };P_{I^{c}}\mathbf{\xi }\right) \psi ^{c}(s,P_{I}\mathbf{\xi })\,\psi
^{c}(s,P_{I^{c}}\mathbf{\xi })\,,  \label{integral}
\end{equation}%
for $n>1$ with $\psi _{0}^{c}(t,\mathbf{\xi })=0$ and $\psi _{1}^{c}(t,%
\mathbf{\xi })=1$ for all $t\in \left[ t_{0},t_{1}\right] $.

The second Ursell function $\psi _{2}^{c}$ satisfies 
\begin{eqnarray}
\psi _{2}^{c}(t,\xi _{1},\xi _{2}) &=&-\int_{t_{0}}^{t}ds\,\sigma _{1}\sigma
_{2}\dot{v}\left( s,x_{1}-x_{2}\right) e^{-\sigma _{1}\sigma _{2}%
\displaystyle\int_{s}^{t}d\tau \,\dot{v}\left( \tau ,x_{1}-x_{2}\right) } 
\notag \\
&=&e^{-\sigma _{1}\sigma _{2}\displaystyle\int_{t_{0}}^{t}d\tau \,\dot{v}%
\left( \tau ,x_{1}-x_{2}\right) }-1~  \label{psi-2}
\end{eqnarray}%
and agrees with (\ref{ursell}) if $\beta \vartheta (\xi _{1},\xi
_{2})=\sigma _{1}\sigma _{2}\left( v\left( t_{1},x_{1}-x_{2}\right) -v\left(
t_{0},x_{1}-x_{2}\right) \right) $. One can verify that there is a unique
solution to the integral equation (\ref{integral}) formally given by the
Ursell functions (\ref{ursell}). These follows from (\ref{diff}) by
induction in $n$. Equation (\ref{integral}), derived for the first time by
Brydges--Kennedy \cite{BK}, is the starting point for the procedure of the
next section.

\section{Majorant Differential Equation\label{MDE}}

\setcounter{equation}{0} \setcounter{theorem}{0}

An one--parameter family $g(t,z)$ of holomorphic function on the open disc 
\begin{equation*}
D_{R}=\left\{ z\in \mathbb{C}:\left\vert z\right\vert <R\right\}
\end{equation*}%
has a convergent power series 
\begin{equation*}
g(t,z)=\sum_{j=0}^{\infty }C_{j}(t)\,z^{j}\text{ }
\end{equation*}%
which we denote by $\mathbf{C}\cdot \mathbf{z}$ with $\mathbf{C}=\left(
C_{j}\right) _{j\geq 0}$ and $\mathbf{z}=\left( z^{j}\right) _{j\geq 0}$.

\begin{definition}
$g(t,z)=\mathbf{C}(t)\cdot \mathbf{z}$ is said to be a uniform majorant of a
function $f(t,z)=\mathbf{c}(t)\cdot \mathbf{z}$ in $\mathcal{D}=\left[ a,b%
\right] \times D_{R}$ if 
\begin{equation*}
\left\vert c_{j}(t)\right\vert \leq C_{j}(t)
\end{equation*}
holds for every $j\in \mathbb{N}$ \ and $t\in \left[ a,b\right] $. We write $%
f\trianglelefteq g$ for the majorant relation.
\end{definition}

Note that a majorant $g$ is a real analytic function and the majorant
relation is preserved by differentiation in $z$ and integration with respect
to $t$.

\smallskip

\noindent \textit{Proof of Theorem \ref{convergence}. Part I.} Our aim is to
find an equation for a uniform majorant $\Phi (t,z)=\mathbf{A}(t)\cdot 
\mathbf{z}$ of the pressure (\ref{p}). Let $\mathbf{a}(t)=\left(
a_{n}(t)\right) _{n\geq 1}$ be given by 
\begin{equation}
a_{n}(t)=\sup_{\xi _{1}\in \Lambda }\int d\rho (\xi _{2})\cdots \int d\rho
(\xi _{n})\,\left\vert \psi _{n}^{c}(t,\mathbf{\xi })\right\vert \,
\label{at}
\end{equation}%
and define 
\begin{equation}
\Gamma (t):=\left\Vert \dot{v}(t)\right\Vert \;,\qquad \qquad \gamma
(s,t):=\int_{s}^{t}\dot{B}(\tau )\,d\tau \,,  \label{Delta}
\end{equation}%
where $B$ and $\left\Vert \cdot \right\Vert $ are given by (\ref{B}) and (%
\ref{norm}). It follows from (\ref{integral}) that 
\begin{equation}
a_{n}(t)\leq \frac{1}{2}\int_{t_{0}}^{t}ds\,e^{n\gamma (s,t)}\Gamma
(s)\sum_{k=1}^{n-1}\binom{n}{k}\,k\,(n-k)\,a_{k}(s)\,a_{n-k}(s)\,~\,~
\label{an}
\end{equation}%
holds for $n>1$ with $a_{1}=1$. We thus have 
\begin{equation}
\left\vert b_{\Lambda ,n}\right\vert \leq \frac{1}{n!}a_{n}(t_{1})\leq
A_{n}(t_{1})  \label{baA}
\end{equation}%
uniformly in $\Lambda $ where $b_{\Lambda ,n}$ is the Mayer coefficients (%
\ref{b}) and $\mathbf{A}(t)=\left( A_{n}(t)\right) _{n\geq 1}$ is a sequence
of continuous functions so that $\left( n!A_{n}(t)\right) _{n\geq 1}$
satisfies (\ref{an}) as an equality: 
\begin{equation}
A_{n}(t)=\frac{1}{2}\int_{t_{0}}^{t}ds\,e^{n\gamma (s,t)}\Gamma
(s)\sum_{k=1}^{n-1}\,k\,A_{k}(s)\,(n-k)\,A_{n-k}(s)\,~  \label{A}
\end{equation}%
for $n>1$ with $A_{1}=1$.

Writing 
\begin{equation}
\Phi \left( t,z\right) :=\sum_{n=1}^{\infty }A_{n}(t)\,z^{n}  \label{Phi0}
\end{equation}%
equations (\ref{p}), (\ref{baA}) and (\ref{A}) yields 
\begin{equation}
\beta p_{\Lambda }(\beta ,z)\trianglelefteq \Phi \left( t_{1},z\right)
\label{pPhi}
\end{equation}%
uniformly in $\Lambda $ where $\Phi \left( t_{1},z\right) $ is the solution
of 
\begin{equation}
\Phi (t,z)=z+\frac{1}{2}\int_{t_{0}}^{t}ds\,\Gamma (s)\left( z\,\frac{%
\partial \Phi }{\partial z}\left( s,z\,e^{\gamma (s,t)}\right) \right) ^{2}
\label{Phi}
\end{equation}%
at time $t=t_{1}$. Here, the following identities have been used 
\begin{eqnarray*}
\sum_{n=2}^{\infty }z^{n}\sum_{k=1}^{n-1}kA_{k}\,(n-k)A_{n-k}
&=&\sum_{n-1=1}^{\infty }\sum_{k=1}^{n-1}kA_{k}z^{k}\,(n-k)A_{n-k}z^{n-k} \\
&=&\sum_{k=1}^{\infty }kA_{k}z^{k}\sum_{n-1=k}^{\infty
}(n-k)A_{n-k}z^{n-k}=\left( z\frac{\partial \Phi }{\partial z}\right) ^{2}\,.
\end{eqnarray*}

The partial derivative of $\Phi $ with respect to $t$ and $z$ are denoted by 
$\Phi _{t}$ and $\Phi _{z}$. From (\ref{Phi0}) and (\ref{Phi}), we have 
\begin{eqnarray*}
\left( z\Phi _{z}\left( s,ze^{\gamma }\right) \right) _{t} &=&\frac{\partial
\gamma }{\partial t}\left( ze^{\gamma }+2^{2}z^{2}e^{2\gamma
}A_{2}(s)+3^{2}z^{3}e^{3\gamma }A_{3}(s)+\cdots \right) \\
&=&\dot{B}ze^{\gamma }\left( z\Phi _{z}\right) _{z}\left( s,ze^{\gamma
}\right) ~,
\end{eqnarray*}%
with $\gamma =\gamma (s,t)$, and 
\begin{equation*}
z\Phi _{z}=z+\int_{t_{0}}^{t}ds\,\Gamma (s)\,z\Phi _{z}\left( s,ze^{\gamma
(s,t)}\right) ~ze^{\gamma (s,t)}\left( z\Phi _{z}\right) _{z}\left(
s,ze^{\gamma (s,t)}\right) ~.
\end{equation*}%
Differentiating (\ref{Phi}) with respect to $t$, 
\begin{equation*}
\Phi _{t}=\frac{1}{2}\Gamma (z\Phi _{z})^{2}+\int_{t_{0}}^{t}ds\,\Gamma
(s)~\,z\Phi _{z}\left( s,ze^{\gamma (s,t)}\right) ~\left( z\Phi _{z}\left(
s,ze^{\gamma (s,t)}\right) \right) _{t}\,
\end{equation*}%
thus yields 
\begin{equation*}
\Phi _{t}=\frac{1}{2}\Gamma (z\Phi _{z})^{2}+\dot{B}\left( z\Phi
_{z}-z\right)
\end{equation*}%
with initial condition $\Phi (t_{0},z)=z$. This establishes the majorant
relation of Theorem \ref{convergence} and equation (\ref{eq}).

\hfill $\Box $

It will be convenient to consider the density 
\begin{equation}
\rho _{\Lambda }(\beta ,z)=z\frac{\partial p_{\Lambda }}{\partial z}(\beta
,z)  \label{rho}
\end{equation}%
whose corresponding Mayer series has the same radius of convergence of the
Mayer series of $p_{\Lambda }$. Writing $\Theta (t,z)=\Phi _{z}(t,z)$, we
have 
\begin{equation}
\frac{\beta }{z}\rho _{\Lambda }(\beta ,z)\trianglelefteq \Theta (\beta ,z)
\label{rhoTheta}
\end{equation}%
with $\Theta $ satisfying the following quasi--linear first order PDE 
\begin{equation}
\Theta _{t}=\Gamma \left( z\Theta ^{2}+z^{2}\Theta \Theta _{z}\right) +\dot{B%
}\left( z\Theta _{z}+\Theta -1\right)  \label{Theta}
\end{equation}%
with initial condition $\Theta (t_{0},z)=1$.

In view of (\ref{Phi0}), (\ref{rho}) and (\ref{rhoTheta}), we seek for
solutions of (\ref{Theta}) of the form%
\begin{equation}
\Theta (t,z)=1+\sum_{n=2}^{\infty }C_{n}(t)z^{n-1}  \label{form}
\end{equation}%
with $C_{n}=nA_{n}(t)\geq 0$ and this form is preserved by the equation.

\begin{remark}
\label{R1} The convergence of Mayer series for an integrable stable
potential ($v(t)$ so that $\displaystyle\int_{t_{0}}^{t_{1}}\left\Vert \dot{v%
}(t)\right\Vert dt<\infty $ and $v(t_{1})=\beta \vartheta $) can be
established as in the case of non negative potential ($\dot{B}\equiv 0$).
Writing $w=e^{\gamma (t_{0},t)}z$ and $\widetilde{A}_{n}=e^{-n\gamma
(t_{0},t)}A_{n}$, we have 
\begin{equation*}
\Phi (t,z)=\sum_{n=1}^{\infty }\widetilde{A}_{n}(t)~w^{n}:=\widetilde{\Phi }%
(t,w)
\end{equation*}%
with $\widetilde{\Phi }$ satisfying $\widetilde{\Phi }_{t}=\left( \Gamma
/2\right) \left( w\widetilde{\Phi }_{w}\right) ^{2}$. The Mayer series of
these systems converges for all $\left( \beta ,z\right) $ so that the
classical solution $\widetilde{\Phi }(t,e^{\gamma (t_{0},t_{1})}|z|)$ exists
up to $t=t_{1}$.
\end{remark}

\section{The Method of Characteristics\label{MC}}

\setcounter{equation}{0} \setcounter{theorem}{0}

In this section the convergence of Mayer series will be addressed from the
point of view of (\ref{Theta}). We seek for a subset of $\left\{ (t,z):t\in %
\left[ t_{0},t_{1}\right] ,z\geq 0\right\} $ in which there exists a unique
classical solution of the initial value problem. We shall distinguish two
cases, depending on whether the stability function $\dot{B}(t)$ is or is not
identically zero.

\subsection{Non Negative Potentials}

Let equation (\ref{Theta}) with $\dot{B}(s)\equiv 0$, 
\begin{equation}
\Theta _{t}=\Gamma \left( z\Theta ^{2}+z^{2}\Theta \Theta _{z}\right) 
\label{Theta2}
\end{equation}%
with $\Theta (t_{0},z)=1$, be addressed by the method of characteristics.
This equation describes, by (\ref{rhoTheta}), a density majorant of a gas
with non negative (repulsive) potential $v\geq 0$.

It is convenient change the $t$ variable by $\tau =\displaystyle%
\int_{t_{0}}^{t}\Gamma (s)~ds$ and write $U(\tau )=\Theta (t(\tau ),z(\tau
)) $. Equation%
\begin{equation*}
U^{\prime }=t^{\prime }\Theta _{t}+z^{\prime }\Theta _{z}
\end{equation*}%
together with (\ref{Theta2}) yield a pair of ordinary differential equations
(ODE) 
\begin{eqnarray}
U^{\prime } &=&zU^{2}  \label{U} \\
z^{\prime } &=&-z^{2}U~  \label{z}
\end{eqnarray}%
satisfying the initial conditions $U(0)=1$ and $z(0)=z_{0}$. Dividing (\ref%
{U}) by (\ref{z}) gives 
\begin{equation*}
\frac{dU}{dz}=-\frac{U}{z}
\end{equation*}%
which can be solved as a function of $z$ 
\begin{equation}
U(z)=\frac{z_{0}}{z}~.  \label{Uz}
\end{equation}%
Inserting (\ref{Uz}) into (\ref{z}), leads to the following characteristic
equation $z^{\prime }=-z_{0}z$ whose solution is 
\begin{equation}
z(\tau )=z_{0}\exp \left\{ -z_{0}\tau \right\} ~.  \label{characteristics}
\end{equation}

Now, $\Theta (t,z)$ is given by (\ref{Uz}) with $z_{0}=z_{0}(\tau ,z)$
implicitly given by (\ref{characteristics}). If $W(x)$ denotes the Lambert $%
W $--function \cite{K}, defined as the inverse of $f(W)=We^{W}$ ($f\left(
W(x)\right) =x$ and $W\left( f(w)\right) =w$), we have 
\begin{equation*}
z_{0}(\tau ,z)=\frac{-1}{\tau }W\left( -z\tau \right) ~
\end{equation*}%
and the solution of the initial value problem (\ref{Theta2}) is thus given by%
\begin{equation*}
\Theta (t,z)=\frac{-1}{z\tau }W\left( -z\tau \right) ~
\end{equation*}%
provided 
\begin{equation}
ez\tau =ez\int_{t_{0}}^{t}\Gamma (s)~ds<1~.  \label{cond1}
\end{equation}%
Since $\Im (W(x))\neq 0$ for $x<-1/e$ in the principal branch of the $W$
--function (see e.g. \cite{K}), this condition has to be imposed to ensure
uniqueness. $W$ has a branching point singularity at $-1/e+i0$ with $%
W(-1/e)=\Re (W(-1/e))=-1$. At the singularity $z^{\ast }=z^{\ast }(t)=\left(
z\tau \right) ^{-1}$, $\Theta (t,z^{\ast })=e$ and the series (\ref{form})
diverges.

\begin{remark}
\label{Rmk3}

\begin{enumerate}
\item For non negative potentials we can actually find a pressure majorant $%
\Phi (\beta ,z)$ explicitly. Differentiating $x=W(x)\exp \left( W(x)\right) $
and solving for $W^{\prime }$ gives, after some manipulations, 
\begin{equation*}
\frac{W}{x}=\left( W+\frac{W^{2}}{2}\right) ^{\prime }
\end{equation*}%
and this, together with (\ref{rho}) and $W(0)=0$, yields 
\begin{equation*}
\Phi (t,z)=\frac{-1}{\tau }\int_{0}^{-z\tau }\frac{W(x)}{x}dx=\frac{-1}{\tau 
}\left( W\left( -z\tau \right) +\frac{1}{2}W^{2}\left( -z\tau \right)
\right) ~.
\end{equation*}

\item The interpretation of (\ref{cond1}) is as follows. Let $\mathcal{T}%
(r,z_{0})=\left( \ln r-\ln z_{0}\right) /\left( r-z_{0}\right) $ be the time
necessary for two characteristics starting from $r$ and $z_{0}$ to meet at a
point $z$: 
\begin{equation*}
z=re^{-r\mathcal{T}}=z_{0}e^{-z_{0}\mathcal{T}}~.
\end{equation*}%
$\mathcal{T}(r,z_{0})$ is a monotone decreasing function of $r$ with $%
\lim_{r\nearrow z_{0}}\mathcal{T}(r,z_{0})=1/z_{0}$. Note that, $\Theta (%
\mathcal{T},z)$ assumes two different values $U(z)=r/z$ and $%
U_{0}(z)=z_{0}/z $ for $r<z_{0}$ which tends to $e$ in the limit $r\nearrow
z_{0}$. The minimum intercepting time of two characteristics with $r\leq
z_{0}$ is attained when both coincide and this occurs exactly at the
branching singularity $-z\mathcal{T}=-1/e$ of $W(x)$ where uniqueness of (%
\ref{Theta2}) breaks down.

\item From condition (\ref{cond1}) and (\ref{pPhi}), the Mayer series (\ref%
{p}) converges if 
\begin{equation}
\left\vert z\right\vert <\left( e\int_{t_{0}}^{t_{1}}\Gamma (s)~ds\right)
^{-1}~.  \label{R}
\end{equation}%
Choosing $v(t)=(t-t_{0})\beta \vartheta /(t_{1}-t_{0})$, we have $%
\displaystyle\int_{t_{0}}^{t_{1}}\Gamma (s)~ds=\beta \left\Vert \vartheta
\right\Vert $ and for stable potentials, according to Remark \ref{R1}, the
Mayer series converges for all $\left( \beta ,z\right) $ so that 
\begin{equation}
e\left\vert z\right\vert \exp \left\{ \int_{t_{0}}^{t_{1}}\dot{B}%
(s)~ds\right\} \int_{t_{0}}^{t_{1}}\Gamma (s)~ds~<1~  \label{RRR}
\end{equation}%
which gives the same condition $e\beta \left\vert z\right\vert e^{\beta
B}\left\Vert \vartheta \right\Vert <1~$obtained by Brydges and Federbush
(see eq. (11) in \cite{BF}).
\end{enumerate}
\end{remark}

\subsection{Stable Potentials\label{SP}}

We now turn to equation (\ref{Theta}). Instead of solving it explicitly, we
shall obtain an upper solution $\overline{\Theta }(t,z)$ whose existence
implies convergence of density majorant series. $\overline{\Theta }(t,z)$
satisfies equation (\ref{Theta2}) for non--negative potential with $\Gamma $
substituted by $\Gamma /f$ for some properly chosen functions $f$. Our
procedure allows us to obtain various convergence conditions depending on
the choice of $f$.

\smallskip

\noindent \textit{Proof of Theorem \ref{convergence}. Conclusion.} Let us
change the variable $z$ by $w=f(t)z$, for a function $f(t)\geq 1$ such that $%
f(0)=1$, and define $\Theta (t,z)=f(t)\Psi \left( t,f(t)z\right) $. Using
the chain rule, we have 
\begin{eqnarray*}
\Theta _{t} &=&f\left( \Psi _{t}+\frac{f^{\prime }}{f}\Psi +\frac{f^{\prime }%
}{f}w\Psi _{w}\right) \\
z\Theta _{z} &=&fw\Psi _{w}
\end{eqnarray*}%
which, together with (\ref{Theta}), yields 
\begin{equation*}
\Psi _{t}-\left( \overset{\cdot }{B}-\frac{f^{\prime }}{f}+\Gamma w\Psi
\right) w\Psi _{w}=\left( \overset{\cdot }{B}-\frac{f^{\prime }}{f}\right)
\Psi +\Gamma w\Psi ^{2}-\frac{\overset{\cdot }{B}}{f}~.~
\end{equation*}%
Choosing 
\begin{equation*}
f(t)=\exp \left\{ \int_{t_{0}}^{t}\dot{B}(s)~ds\right\}
\end{equation*}%
yields%
\begin{equation}
\Psi _{t}-\Gamma w^{2}\Psi \Psi _{w}=\Gamma w\Psi ^{2}-\frac{\dot{B}}{f}\leq
\Gamma w\Psi ^{2}  \label{Psi}
\end{equation}%
by positivity of $f$ and $\dot{B}$. Note that $\Psi $ satisfies (\ref{Theta2}%
) as an inequality.

As \ before, we define $\tau =\displaystyle\int_{t_{0}}^{t}\Gamma (s)~ds$
and write

\begin{equation}
U(\tau )=\Psi (t(\tau ),w(\tau )).  \label{Utau}
\end{equation}%
From $U^{\prime }=\Psi _{t}t^{\prime }+\Psi _{w}w^{\prime }$and (\ref{Psi}),
we have 
\begin{eqnarray}
U^{\prime } &\leq &wU^{2}  \label{U1} \\
w^{\prime } &=&-w^{2}U  \label{z1}
\end{eqnarray}%
with initial conditions $U(0)=1$ and $w(0)=w_{0}$. Equations (\ref{U1}) and (%
\ref{z1}) can be written in terms of the variables $V(\tau )=wU$ and $w$ as 
\begin{eqnarray}
V^{\prime } &\leq &0  \notag \\
\frac{w^{\prime }}{w} &=&-V  \label{w/w}
\end{eqnarray}%
with $V(0)=w_{0}$, which implies 
\begin{equation}
V(\tau )\leq w_{0}  \label{VV}
\end{equation}%
and 
\begin{equation}
w(\tau )\geq w_{0}\exp \left\{ -w_{0}\tau \right\} ~.  \label{ineq}
\end{equation}%
Here, we need an inequality for $w_{0}=w_{0}(\tau ,z)$ so that (\ref{ineq})
holds. Writing (\ref{ineq}) as 
\begin{equation*}
-w_{0}\tau ~e^{-w_{0}\tau }\geq -w\tau ~,
\end{equation*}%
we have 
\begin{equation}
w_{0}(\tau ,w)\leq -\frac{1}{\tau }W\left( -w\tau \right) ~,  \label{w0}
\end{equation}%
where $W$ is the Lambert $W$--function, provided%
\begin{equation*}
w\tau ~=z\exp \left\{ \int_{t_{0}}^{t}\dot{B}(s^{\prime })~ds^{\prime
}\right\} \int_{t_{0}}^{t}\Gamma (s)~ds~<1/e~.
\end{equation*}%
Plugging (\ref{w0}) into (\ref{VV}), together with (\ref{Utau}), leads to 
\begin{equation}
\Theta (t,z)=\frac{V(\tau (t))}{z}\leq \overline{\Theta }(t,z)~,
\label{Thetabar}
\end{equation}%
where 
\begin{equation*}
\overline{\Theta }(t,z)=\frac{-1}{z\displaystyle\int_{t_{0}}^{t}\Gamma (s)~ds%
}W\left( -z\exp \left\{ \int_{t_{0}}^{t}\dot{B}(s)~ds\right\}
\int_{t_{o}}^{t}\Gamma (s)~ds~\right) ~,
\end{equation*}%
recovering the statements of Remark \ref{R1}.

We now go beyond inequality (\ref{ineq}). For this we modify slightly the
previous scaling by defining $\Theta (t,z)=\Psi ^{(1)}\left(
t,f_{1}(t)z\right) $. By the chain rule, $\Psi ^{(1)}$ is the solution of 
\begin{equation*}
\Psi _{t}-\frac{\Gamma }{f_{1}}w^{2}\Psi \Psi _{w}=\left( \dot{B}-\frac{%
f_{1}^{\prime }}{f_{1}}\right) w\Psi _{w}+\dot{B}\Psi -\dot{B}+\frac{\Gamma 
}{f_{1}}w\Psi ^{2}~~
\end{equation*}%
satisfying $\Psi ^{(1)}(t_{0},w)=1$. Choosing 
\begin{equation}
f_{1}(t)=\exp \left\{ 2\int_{t_{0}}^{t}\dot{B}(s)~ds\right\}  \label{f}
\end{equation}%
and using 
\begin{eqnarray*}
-w\Psi _{w}^{(1)}+\Psi ^{(1)}-1 &=&-\sum_{n=2}^{\infty
}(n-1)C_{n}^{(1)}w^{n-1}+\sum_{n=1}^{\infty }C_{n}^{(1)}w^{n-1}-1 \\
&=&-\sum_{n=2}^{\infty }(n-2)C_{n}^{(1)}w^{n-1}\leq 0
\end{eqnarray*}%
in view of (\ref{form}) and $w=zf_{1}(t)\geq 0$, we have%
\begin{equation}
\Psi _{t}-\frac{\Gamma }{f_{1}}w^{2}\Psi \Psi _{w}=\dot{B}\left( -w\Psi
_{w}+\Psi -1\right) +\frac{\Gamma }{f_{1}}w\Psi ^{2}\leq \frac{\Gamma }{f_{1}%
}w\Psi ^{2}~.  \label{Psi1}
\end{equation}

We define $\tau =\displaystyle\int_{t_{0}}^{t}\Gamma (s)/f_{1}(s)~ds$ and
repeat the steps from (\ref{Utau}) to (\ref{Thetabar}). Provided 
\begin{equation}
w\tau ~=z\tau _{1}<1/e  \label{wtau}
\end{equation}%
holds with $\tau _{1}$ given by 
\begin{equation}
\tau _{1}(t)~=\int_{t_{0}}^{t}e^{2\displaystyle\int_{s}^{t}\dot{B}%
(s)~ds}\Gamma (s)~ds~,~  \label{BK}
\end{equation}%
the solution of the initial value problem (\ref{Theta}) satisfies 
\begin{equation}
\Theta (t,z)\leq ~\frac{-1}{\tau _{1}z}W\left( -\tau _{1}z\right) ~.
\label{W}
\end{equation}

It follows from (\ref{Thetabar}), (\ref{W}) and Remark \ref{Rmk3}.$1$. that (%
\ref{Phitz}) holds and converges either under the condition (\ref{RRR}) or
under (\ref{wtau}). This concludes the proof of Theorem \ref{convergence}.

\hfill $\Box $

\begin{remark}
\label{R2}

\begin{enumerate}
\item Working directly with the integral equation (\ref{integral}),
Brydges--Kennedy \cite{BK1} have established recursively 
\begin{equation*}
a_{n}(t)\leq \tau _{1}{}^{n-1}n^{n-2}
\end{equation*}%
with $\tau _{1}$ and $a_{n}$ given by (\ref{BK}) and (\ref{an}). Hence, the
density majorant of stable potentials satisfies 
\begin{eqnarray*}
\Theta (t,z) &=&\sum_{n=1}^{\infty }\frac{a_{n}(t)}{(n-1)!}z^{n-1} \\
&\leq &\frac{-1}{\tau _{1}z}\sum_{n=1}^{\infty }\frac{(-n)^{n-1}}{n!}\left(
-\tau _{1}z\right) ^{n}~,
\end{eqnarray*}%
and this, by the Taylor series of the Lambert $W$--function \cite{K}, is
exactly (\ref{W}).

\item \label{Rmk2}Criterion (\ref{wtau}) for convergence may be more
advantageous than (\ref{RRR}) when $v(t)$ is a multiscale decomposition of a
potential. As observed by Brydges--Kennedy (see Remarks at the end of \cite%
{BK1}), a poor stability bound (large $\dot{B}(t)$) at short distances ($%
t\ll 0$) may be overcome by the smallness of the interaction $\Gamma
=\left\Vert \dot{v}(t)\right\Vert $ at these scales.
\end{enumerate}
\end{remark}

\section{Application to Yukawa Gas\label{AA}}

\setcounter{equation}{0} \setcounter{theorem}{0}

The majorant method is applied to investigate whether the Mayer series for
the Yukawa gas is convergent in the region of collapse. The first and third
subsections redo the proofs in \cite{B,BK} and \cite{BK} for $\beta <4\pi $
and $4\pi \leq \beta <16\pi /3$, respectively. After introducing the model
we illustrate how useful is condition (\ref{wtau}) for $\beta <4\pi $. The
procedure of Subsection \ref{SP} is then modified in order to overcome the
first threshold at $\beta _{1}=4\pi $. In the present section we intend to
show the capabilities of our majorant method to extend results in \cite{BK}.
The proof of Theorem \ref{singularity} will be deferred to Section \ref{Cjt}.

\subsection{Convergence of the Mayer Series for $\protect\beta <4\protect\pi 
$}

Particles in this system have assigned two charges $\sigma \in \left\{
-1,1\right\} $ and they interact via the Yukawa potential%
\begin{eqnarray}
\vartheta (\xi _{1},\xi _{2}) &=&\sigma _{1}\sigma _{2}\left( -\Delta
+1\right) ^{-1}(x_{1}-x_{2})  \notag \\
&=&\frac{\sigma _{1}\sigma _{2}}{\left( 2\pi \right) ^{2}}\int_{\mathbb{R}%
^{2}}\frac{e^{-ik\cdot (x_{1}-x_{2})}}{k^{2}+1}d^{2}k  \label{Yp}
\end{eqnarray}%
where $-\Delta $ is the Laplace operator in $\mathbb{R}^{2}$. Potential (\ref%
{Yp}) violates condition $1.$ of (\ref{w}), since 
\begin{equation*}
\left\vert \vartheta (\xi ,\xi )\right\vert =\frac{1}{4\pi }%
\lim_{R\rightarrow \infty }\int_{0}^{R}\frac{2\eta }{\eta ^{2}+1}d\eta =%
\frac{1}{4\pi }\lim_{R\rightarrow \infty }\ln \left( R^{2}+1\right)
\end{equation*}%
diverges logarithmically. We replace $\beta \vartheta (\xi _{1},\xi _{2})$
by $\sigma _{1}\sigma _{2}v(t,x_{1}-x_{2})$ given by 
\begin{equation}
v(t,x)=\beta \left( -\Delta +\kappa (t)\right) ^{-1}(x)=\frac{\beta }{2\pi }%
~K_{0}\left( \sqrt{\kappa (t)}\left\vert x\right\vert \right)  \label{C}
\end{equation}%
where $K_{0}$ is the modified Bessel function of second kind (see e.g. \cite%
{GJ} p. $126$) and $\kappa $ is a monotone decreasing scale function such
that 
\begin{equation*}
\kappa (-\infty )=\infty \qquad \mathrm{and}\qquad \kappa (0)=1~,
\end{equation*}%
and decompose it into scales 
\begin{equation*}
v(t,x)=\int_{-\infty }^{t}\dot{v}(s,x)~ds
\end{equation*}%
where 
\begin{equation}
\dot{v}(s,x)=\frac{\beta }{4\pi }\frac{\mathbb{\kappa }^{\prime }(s)}{\sqrt{%
\kappa (s)}}\left\vert x\right\vert K_{0}^{\prime }\left( \sqrt{\kappa (s)}%
\left\vert x\right\vert \right) ~.  \label{Cdot}
\end{equation}%
By definitions (\ref{B}) and (\ref{Delta}), we have

\begin{proposition}
For each $-\infty <s<0$, $\dot{v}(s,x-y)$ is a positive kernel satisfying
properties $1-3$ of (\ref{w}) with 
\begin{equation}
\dot{B}(s)=\frac{1}{2}\left\vert \dot{v}(s,0)\right\vert =\frac{\beta }{8\pi 
}\left( -\ln ~\kappa (s)\right) ^{\prime }~  \label{B2}
\end{equation}%
and%
\begin{equation}
\Gamma (s)=\Vert \dot{v}(s,\cdot )\Vert =\beta ~\frac{-\mathbb{\kappa }%
^{\prime }(s)}{\kappa ^{2}(s)}.  \label{Delta2}
\end{equation}
\end{proposition}

\noindent \textit{Proof.} Properties $1-3$ of (\ref{w}) follow immediately
from the expressions (\ref{C}) and (\ref{Yp}), provided (\ref{B2}) and (\ref%
{Delta2}) are established. Applying Cauchy formula for the integral (\ref{Yp}%
) (with $1$ replaced by $\kappa (t)$) in the $x-y$ direction, gives 
\begin{equation*}
v(t,x-y)=\frac{\beta }{2\pi }g\left( \sqrt{\kappa (t)}\left\vert
x-y\right\vert \right)
\end{equation*}%
with 
\begin{equation}
g(w):=\int_{0}^{\infty }\frac{e^{-w\sqrt{k^{2}+1}}}{\sqrt{k^{2}+1}}dk~.~
\label{g}
\end{equation}%
By the chain rule%
\begin{equation}
\left( g\circ w\right) ^{\prime }=-w^{\prime }\int_{0}^{\infty }e^{-w\sqrt{%
k^{2}+1}}dk~=-\frac{w^{\prime }}{w}\int_{w}^{\infty }\frac{e^{-\zeta }}{%
\sqrt{\zeta ^{2}-w^{2}}}\zeta d\zeta \equiv -\frac{w^{\prime }}{w}h(w)
\label{gw}
\end{equation}%
and equation (\ref{Cdot}) can thus be written as%
\begin{equation}
\dot{v}(s,x-y)=\frac{-\beta }{4\pi }\frac{\mathbb{\kappa }^{\prime }(s)}{%
\kappa (s)}h\left( \sqrt{\kappa (s)}~\left\vert x-y\right\vert \right) ~,
\label{omega}
\end{equation}%
which gives (\ref{B2}) by taking $\left\vert x-y\right\vert =0$. Note that $
- \kappa ^{\prime }(s)\geq 0$, by assumption. Integrating (\ref{g}%
) times $w$, yields 
\begin{equation}
\int_{0}^{\infty }g(w)wdw=\int_{0}^{\infty }\frac{1}{\left( k^{2}+1\right)
^{3/2}}dk=1  \label{gg}
\end{equation}%
which implies 
\begin{eqnarray*}
\Vert \dot{v}(s,\cdot )\Vert &=&\frac{\beta }{2\pi }\frac{d}{ds}%
\int_{0}^{\infty }g\left( \sqrt{\kappa (s)}r\right) 2\pi rdr \\
&=&\beta \left( \frac{1}{\kappa (s)}\right) ^{\prime }
\end{eqnarray*}%
and concludes the proof.

\hfill $\Box $

The majorant density $\Theta (t,z)$ of the Yukawa gas satisfies equation (%
\ref{Theta}) with $\dot{B}$ and $\Gamma $ given by (\ref{B2}) and (\ref%
{Delta2}), respectively, and initial data $\Theta (t_{0},z)=1$ where $t_{0}$
plays the role of a short distance cutoff, which is removed if $%
t_{0}\rightarrow $ $-\infty $.

By (\ref{rhoTheta}) and Remark \ref{R2}.\ref{Rmk2}, the Mayer series
converges if (\ref{wtau}) holds. We have 
\begin{equation*}
\lim_{t_{0}\rightarrow -\infty }\tau _{1}(0)=-\beta \lim_{t_{0}\rightarrow
-\infty }\int_{t_{0}}^{0}\frac{\kappa ^{\prime }(s)}{\kappa
^{2-\beta /4\pi }(s)}ds=\frac{4\pi \beta }{4\pi -\beta }\lim_{t_{0}%
\rightarrow -\infty }\left\{ 1-\kappa ^{\beta /4\pi -1}(t_{0})\right\} =%
\frac{4\pi \beta }{4\pi -\beta }
\end{equation*}%
provided $\beta <4\pi $, independently of the scale function $\kappa $. The
Mayer series of Yukawa gas converges if 
\begin{equation*}
\beta <4\pi \qquad \mathrm{and}\qquad \left\vert z\right\vert <\frac{4\pi
-\beta }{4\pi e\beta }~.
\end{equation*}%
Although it has been previously established in \cite{B,BK}, our method is
independent of the scale function $\kappa $ and improves their radius of
convergence.

\subsection{The $2$--Regularized Majorant Solution\label{2-R}}

The Mayer series for the Yukawa gas between the first and second thresholds, 
$4\pi \leq \beta <6\pi $, diverges because the norm $a_{2}(t)$ of second
Ursell function $\psi _{2}^{c}(t,\mathbf{\xi })$ diverges when the
ultraviolet cutoff is removed (see e.g. \cite{BGN, BK} and references
therein). We are going to show that the regularized density majorant $\Theta
^{(2)}(t,z)$, defined by $\Theta (t,z)$ with the singular part of $a_{2}(t)$
removed by a Lagrange multiplier, exists and implies convergence of the
Mayer series for $\beta <16\pi /3$. For $16\pi /3\leq \beta <6\pi $, the
upper bound $A_{3}$ for the norm of $\psi _{3}^{c}(t,\mathbf{\xi })$ also
diverges and equation (\ref{an}) with $n=3$ has to be modified in order the
majorant relation (\ref{rhoTheta}) holds with $\Theta $ replaced by $\Theta
^{(2)}$. The modified equation requires an improved stability condition
which prevents $\beta _{2}=16\pi /3$, and more generally $\beta =\beta _{2r}$%
, $r=1,2,\ldots $, where 
\begin{equation}
\beta _{n}=8\pi \left( 1+\frac{1}{n}\right) ^{-1}=8\pi \frac{n}{n+1}~,
\label{betan}
\end{equation}%
to be considered a threshold of the Yukawa model (see Remarks on pg. $47$ of 
\cite{BK}).

The second Ursell function $\psi _{2}^{c}(t,\mathbf{\xi })$ is explicitly
given by (\ref{psi-2}). Together with (\ref{C}), we have 
\begin{equation}
\psi _{2}^{c}(t,\xi _{1},\xi _{2})=\exp \left\{ \frac{-\beta \sigma
_{1}\sigma _{2}}{2\pi }~\left( K_{0}\left( \sqrt{\kappa (t)}r\right)
-K_{0}\left( \sqrt{\kappa (t_{0})}r\right) \right) \right\} -1  \label{psi2}
\end{equation}%
where $r=\left\vert x_{1}-x_{2}\right\vert $.

To isolate the singularity of (\ref{psi2}) we replace $a_{2}(t)=\sup_{\xi
_{1}\in \Lambda }\displaystyle\int d\rho (\xi _{2})\left\vert \psi
_{2}^{c}(t,\xi _{1},\xi _{2})\right\vert $ by 
\begin{equation*}
a_{2}^{\pm }(t)=\sup_{x_{1}}\int d^{2}x_{2}\frac{1}{2}\sum_{\substack{ %
\sigma _{1},\sigma _{2}\in \left\{ -1,1\right\} :  \\ \sigma _{1}\sigma
_{2}=\pm }}\left\vert \psi _{2}^{c}(t,\xi _{1},\xi _{2})\right\vert
\end{equation*}%
which distinguish whether the charge product $\sigma _{1}\sigma _{2}$ is
positive or negative ($\sigma _{1}\sigma _{2}=-1\Leftrightarrow \sigma
_{1}+\sigma _{2}=0$). Using the fact that $K_{0}(x)$ is a monotone
decreasing function together with $1-e^{-y}\leq y$ for $y\geq 0$, we have 
\begin{eqnarray*}
a_{2}^{+}(t) &\leq &\beta \int_{0}^{\infty }r\left( K_{0}\left( \sqrt{\kappa
(t)}r\right) dr-K_{0}\left( \sqrt{\kappa (t_{0})}r\right) \right) dr \\
&\leq &\beta \left( \frac{1}{\kappa (t)}-\frac{1}{\kappa (t_{0})}\right) ~.
\end{eqnarray*}%
For $\sigma _{1}\sigma _{2}=-1$, the first line of (\ref{psi-2}), $\dot{v}%
(s,r)\geq 0$ and (\ref{omega}), imply 
\begin{eqnarray}
a_{2}^{-}(t) &=&\frac{\beta }{2}\int_{0}^{\infty }dw~w~h(w)\int_{t_{0}}^{t}%
\frac{-\mathbb{\kappa }^{\prime }(s)}{\kappa ^{2}(s)}\exp \left\{ \frac{%
\beta }{4\pi }\int_{s}^{t}\frac{-\mathbb{\kappa }^{\prime }(\tau )}{\kappa
(\tau )}h\left( \sqrt{\kappa (\tau )/\kappa (s)}w\right) d\tau \right\} ~ds
\label{a2-} \\
&\leq &\frac{2\pi \beta }{4\pi -\beta }\int_{0}^{\infty }dw~w~h(w)\left( 
\frac{1}{\kappa (t)}-\frac{1}{\kappa (t_{0})}\left( \frac{\kappa (t_{0})}{%
\kappa (t)}\right) ^{\beta /4\pi }\right)  \notag \\
&=&\frac{4\pi \beta }{4\pi -\beta }\left( \frac{1}{\kappa (t)}-\frac{1}{%
\kappa (t_{0})}\left( \frac{\kappa (t_{0})}{\kappa (t)}\right) ^{\beta /4\pi
}\right) ~.  \notag
\end{eqnarray}%
Note that $h(w)$ is a monotone decreasing function with $h(0)=1$ and 
\begin{equation*}
\int_{0}^{\infty }dw~w~h(w)=\int_{0}^{\infty }dw~w^{2}~g^{\prime
}(w)=2\int_{0}^{\infty }dw~w~g(w)=2
\end{equation*}%
by (\ref{gw}), integration by parts and (\ref{gg}). As a consequence, both $%
a_{2}^{+}$ and $a_{2}^{-}$ converges as $t_{0}\rightarrow \infty $ if $\beta
<4\pi $ ($a_{2}^{+}$ remains bounded for any $\beta >0$). If $\beta >4\pi $,
since the main contribution of (\ref{a2-}) comes from a neighborhood of $w=0$%
, there exist $\varepsilon >0$ and a constant $C=C(\varepsilon )>0$ such
that $\dfrac{\beta }{4\pi }-\varepsilon >1$ and 
\begin{equation}
a_{2}^{-}(t)\geq C\left( \frac{1}{\kappa (t_{0})}\left( \frac{\kappa (t_{0})%
}{\kappa (t)}\right) ^{\beta /4\pi -\varepsilon }-\frac{1}{\kappa (t)}\right)
\label{a-2}
\end{equation}%
diverges as $t_{0}\rightarrow -\infty $.

Now we shall write the majorant equation (\ref{Theta}) with the singularity
of $A_{2}$ removed by a Lagrange multiplier. This can be done in several
ways. We may define $\Theta ^{R}$ such that $\Theta _{z}^{R}(t,0)=0$ for all 
$t\geq 0$, as the solution of 
\begin{equation*}
\Theta _{t}=\Gamma \left( z\Theta ^{2}+z^{2}\Theta \Theta _{z}\right) +\dot{B%
}\left( z\Theta _{z}+\Theta -1\right) -z\left( \Gamma +2\dot{B}\Theta
_{z}(t,0)\right) ~
\end{equation*}%
with $\Theta (0,z)=1$. Note that $d\Theta _{z}^{R}(t,0)/dt=\Theta
_{tz}^{R}(t,0)=0$ holds for all $t\geq 0$. Another way adopt here is as
follows. The linear part of equation (\ref{Theta}) is responsible for the
singularity of $A_{2}$ if $\beta >4\pi $ and we may choose, instead, a
Lagrange multiplier proportional to $\dot{B}$%
\begin{equation}
\Theta _{t}=\Gamma \left( z\Theta ^{2}+z^{2}\Theta \Theta _{z}\right) +\dot{B%
}\left( z\Theta _{z}+\Theta -1\right) -\frac{1}{2}z\dot{B}\Theta
_{z}(t,0)~~~~~  \label{ThetaR}
\end{equation}%
(with a quarter of intensity needed to make $A_{2}$ vanishes).

Let us denote the solution of (\ref{ThetaR}) by $\Theta ^{(2)}$. As in
Subsection \ref{SP}, if $\Theta ^{(2)}(t,z)=\Psi ^{(2)}(t,w)$, where $%
w=f_{2}(t)z$ and 
\begin{equation}
f_{2}(t)=\exp \left\{ \frac{3}{2}\int_{t_{0}}^{t}\dot{B}(s)~ds\right\} ~,~
\label{ff}
\end{equation}%
using (with $\Psi ^{(2)}=1+\displaystyle\sum_{n=2}^{\infty
}C_{n}^{(2)}w^{n-1}$, $C_{n}^{(2)}\geq 0$)%
\begin{eqnarray*}
-\frac{1}{2}w\Psi _{w}^{(2)}+\Psi ^{(2)}-1-\frac{1}{2}w\left( \Psi
_{w}^{(2)}(t,0)\right) &=&-\frac{1}{2}\sum_{n=2}^{\infty
}(n-1)C_{n}^{(2)}w^{n-1}+\sum_{n=1}^{\infty }C_{n}^{(2)}w^{n-1}-1-\frac{1}{2}%
C_{2}w \\
&=&-\frac{1}{2}\sum_{n=3}^{\infty }(n-3)C_{n}^{(2)}w^{n-1}\leq 0
\end{eqnarray*}%
it follows that $\Psi ^{(2)}$ is the solution of 
\begin{eqnarray*}
\Psi _{t}-\frac{\Gamma }{f_{2}}w^{2}\Psi \Psi _{w} &=&\left( \dot{B}-\frac{%
f_{2}^{\prime }}{f_{2}}\right) w\Psi _{w}+\dot{B}\left( \Psi -1-\frac{1}{2}%
C_{2}w\right) +\frac{\Gamma }{f_{2}}w\Psi ^{2} \\
&=&\dot{B}\left( -\frac{1}{2}w\Psi _{w}+\Psi -1-\frac{1}{2}C_{2}w\right) +%
\frac{\Gamma }{f_{2}}w\Psi ^{2} \\
&\leq &\frac{\Gamma }{f_{2}}w\Psi ^{2}~
\end{eqnarray*}%
satisfying $\Psi ^{(2)}(t_{0},w)=1$. Repeating the steps from (\ref{Utau})
to (\ref{Thetabar}) with $\tau =\dint_{t_{0}}^{t}(\Gamma /f_{2})(s)~ds$,
provided $w\tau =z\tau _{2}<1/e$ where 
\begin{equation}
\tau _{2}~(t)~=\int_{t_{0}}^{t}\exp \left\{ \frac{3}{2}\int_{s}^{t}\dot{B}%
(s^{\prime })~ds^{\prime }\right\} ~\Gamma (s)~ds,  \label{BK1}
\end{equation}%
the solution $\Theta ^{(2)}$ of the initial value problem (\ref{ThetaR})
satisfies%
\begin{equation}
\Theta ^{(2)}(t,z)\leq \frac{-1}{\tau _{2}z}W\left( -\tau _{2}z\right) ~.
\label{W1}
\end{equation}%
Substituting (\ref{B2}) and (\ref{Delta2}) into (\ref{BK1}) with $t=0$, we
have%
\begin{equation*}
\lim_{t_{0}\rightarrow -\infty }\tau _{2}~(0)=\beta \lim_{t_{0}\rightarrow
-\infty }\int_{t_{0}}^{0}\frac{-\mathbb{\kappa }^{\prime }(s)}{\kappa
^{2-3\beta /16\pi }(s)}ds=\frac{16\pi \beta }{16\pi -3\beta }
\end{equation*}%
independently of the scale function $\kappa $, and the power series 
\begin{equation*}
\Theta ^{(2)}(t,z)=1+\sum_{n=2}^{\infty }C_{n}(t)~z^{n-1}
\end{equation*}%
converges if $\beta <16\pi /3$ and $\left\vert z\right\vert <\left( e\tau
_{2}\right) ^{-1}$.

\subsection{Convergence of the Mayer Series for $4\protect\pi \leq \protect%
\beta <16\protect\pi /3$}

$\Theta ^{(2)}(0,z)$ is a candidate to be a majorant of the density $\beta
\rho (\beta ,z)/z$ with the $O\left( z\right) $ term omitted (see equation (%
\ref{rhoTheta})). Its linear coefficient $C_{2}$, which satisfies%
\begin{equation}
\dot{C}_{2}\leq \frac{3}{2}\dot{B}C_{2}+\Gamma ~,\qquad ~\qquad t_{0}<t\leq 0
\label{C2}
\end{equation}%
with $C_{2}(t_{0})=0$, is positive and bounded by $16\pi \beta /(16\pi
-3\beta )$, in view of the fact $\tau _{2}$, defined by (\ref{BK1}), solves (%
\ref{C2}) as an equality. Hence, by (\ref{p}) and (\ref{rho}), the Mayer
series of Yukawa gas with the leading term removed converges uniformly in $%
t_{0}$ for $4\pi \leq \beta <16\pi /3$ if 
\begin{equation}
nb_{n}=n\lim_{\Lambda \nearrow \mathbb{R}^{2}}b_{n,\Lambda }\leq C_{n}
\label{nbnCn}
\end{equation}%
is shown to hold for all $n>2$. The inequality (\ref{nbnCn}), however, does
not follow from (\ref{an}) because the r.h.s. of that equation depends on $%
a_{2}(t)=a_{2}^{+}(t)/2+a_{2}^{-}(t)/2$ which, as shown in Subsection \ref%
{2-R}, diverges as $t_{0}\rightarrow \infty $. We are going to show that $%
a_{2}(t)$ in the r.h.s. of (\ref{an}) can indeed be replaced by $%
a_{2}^{+}(t) $ and this implies (\ref{nbnCn}).

Define

\begin{equation}
F(t,P_{n}\mathbf{\xi })=e^{-\displaystyle\int_{s}^{t}d\tau \,\dot{U}%
_{n}(\tau ,\mathbf{\xi })}\dot{U}\left( s,P_{I}\mathbf{\xi };P_{I^{c}}%
\mathbf{\xi }\right) \psi ^{c}(s,P_{I}\mathbf{\xi })\,\,,  \label{F}
\end{equation}%
and suppose $n\geq 3$ and $\left\vert I\right\vert =2$. Using the stability
bound (\ref{B}) together with (\ref{at}) and (\ref{Delta}), we have%
\begin{eqnarray}
\prod_{l\in I}\int d\rho (\xi _{l})\left\vert F(t,P_{n}\mathbf{\xi }%
)\right\vert &\leq &e^{n\gamma (s,t)}\sum_{i\in I^{c}}\left\vert \sigma
_{i}\right\vert \prod_{l\in I}\int d\rho (\xi _{l})\left\vert \sum_{j\in I}%
\dot{v}(s,x_{i}-x_{j})\sigma _{j}\psi ^{c}(s,P_{I}\mathbf{\xi })\,\right\vert
\notag \\
&\leq &e^{n\gamma (s,t)}\Gamma (s)\left\vert I^{c}\right\vert \sup_{x}\int
d^{2}x^{\prime }\frac{1}{4}\sum_{\sigma ,\sigma ^{\prime }\in \left\{
-1,1\right\} }\left\vert \sigma +\sigma ^{\prime }\right\vert \left\vert
\psi _{2}^{c}(s,\xi ,\xi ^{\prime })\right\vert  \notag \\
&=&e^{n\gamma (s,t)}\Gamma (s)\left\vert I^{c}\right\vert ~a_{2}^{+}(s)
\label{mod}
\end{eqnarray}%
by translational invariance. Since $a_{2}^{+}$ satisfies the equation%
\begin{equation*}
\dot{a}_{2}^{+}\leq \frac{\Gamma }{2}~,\qquad ~\qquad t_{0}<t\leq 0~,
\end{equation*}%
with $a_{2}^{+}(t_{0})=0$ (for $n=2$ the stability condition (\ref{stb1})
holds with $\delta _{2}=2$), $a_{2}^{+}(t)<A_{2}(t)=C_{2}(t)/2$ holds for
all $t\in \left[ t_{0},0\right] $. Hence, including the modification (\ref%
{mod}), equation (\ref{A}) majorates (\ref{an}) and (\ref{nbnCn}) holds in
view of (\ref{baA}).

The Mayer series of Yukawa gas, with the singular part of the norm of $\psi
_{2}^{c}$ removed, converges if 
\begin{equation*}
\beta <\frac{16}{3}\pi \qquad \mathrm{and}\qquad \left\vert z\right\vert <%
\frac{16\pi -3\beta }{16\pi \beta e}~
\end{equation*}%
and this establishes the conjecture up to $\beta _{2}.$

\hfill $\Box $

\begin{remark}
The authors of \cite{BK} have established convergence of $\Theta $ for $%
\beta <16\pi /3$ with $O\left( z\right) $ term omitted directly from the
integral equation (\ref{integral}). Our method employs the same ingredients
but the invariance under translation is used right in the beginning, and not
after various manipulations of equation (\ref{mod}). This little detail
improves the radius of convergence (see Theorem $4.3$ of \cite{BK}) and
allows us to go beyond the second threshold.
\end{remark}

\section{Proof of BGN's Conjecture\label{Cjt}}

The procedure of Section \ref{AA} will now be extended in order to prove
Theorem \ref{singularity}. Our proof generalizes and includes all results
obtained previously.

The regularized density majorant $\Theta ^{(k)}(\beta ,z)$, defined by $%
\Theta (\beta ,z)$ with the singular part removed by a Lagrange multipliers,
is introduced as follows. If $\beta <\beta _{k}=8\pi \left( 1+k^{-1}\right)
^{-1}$ the $k$--regularized majorant is the solution of equation 
\begin{equation}
\Theta _{t}=\Gamma \left( z\Theta ^{2}+z^{2}\Theta \Theta _{z}\right) +\dot{B%
}\left( z\Theta _{z}+\Theta -L_{k}\right)  \label{ThetaRn}
\end{equation}%
where%
\begin{equation*}
L_{k}=L_{k}(t)=1+\sum_{j=1}^{k-1}\frac{1}{j!}\left( 1-\frac{j}{k}\right)
z^{j}\left( \Theta _{\underset{j-\mathrm{times}}{\underbrace{z\cdots z}}%
}(t,0)\right)
\end{equation*}%
with $\Theta (0,z)=1$. We analogously define $\Theta ^{(k)}(t,z)=\Psi
^{(k)}(t,w)$, where $w=f_{k}(t)z$ and 
\begin{equation}
f_{k}(t)=\exp \left\{ \frac{k+1}{k}\int_{t_{0}}^{t}\dot{B}(s)~ds\right\} ~.~
\label{fn}
\end{equation}%
Using 
\begin{equation*}
\left( \dot{B}-\dfrac{f_{k}^{\prime }}{f_{k}}\right) w\Psi _{w}^{(k)}+\dot{B}%
\left( \Psi ^{(k)}-1\right) -\dot{B}\sum_{m=2}^{k}\left( 1-\dfrac{m-1}{k}%
\right) C_{m}^{(k)}w^{m-1}
\end{equation*}%
\begin{equation*}
=-\dfrac{1}{k}\sum_{m=k+1}^{\infty }(m-k-1)C_{m}^{(k)}w^{m-1}\leq 0
\end{equation*}%
together with (\ref{ThetaRn}), $\Psi ^{(k)}$ is the solution of 
\begin{equation*}
\Psi _{t}-\frac{\Gamma }{f_{k}}w^{2}\Psi \Psi _{w}\leq \frac{\Gamma }{f_{k}}%
w\Psi ^{2}~
\end{equation*}%
satisfying $\Psi ^{(k)}(0,w)=1$. Repeating the steps from (\ref{Utau}) to (%
\ref{Thetabar}), the solution $\Theta ^{(k)}$ of the initial value problem (%
\ref{ThetaRn}) satisfies 
\begin{equation}
\Theta ^{(k)}(t,z)\leq \frac{-1}{\tau _{k}z}W\left( -\tau _{k}z\right) ~
\label{Thetan}
\end{equation}%
provided $\tau _{k}z<1/e$ where, by (\ref{betan}), 
\begin{equation}
\tau _{k}=\int_{t_{0}}^{t}\exp \left\{ \frac{k+1}{k}\int_{s}^{t}\dot{B}%
(s^{\prime })~ds^{\prime }\right\} ~\Gamma (s)~ds~=\frac{\beta \beta _{k}}{%
\beta _{k}-\beta }\left( \frac{1}{\kappa (t)}-\frac{1}{\kappa (t_{0})}\left( 
\frac{\kappa (t_{0})}{\kappa (t)}\right) ^{\beta /\beta _{k}}\right)
\label{taun}
\end{equation}%
Hence, (\ref{Thetan}) defines a convergent power series in $z$ , uniformly
in $t_{0}$, if $\beta <\beta _{k}$ and $\left\vert z\right\vert <\left(
e\tau _{k}\right) ^{-1}$.

We now examine whether $\Theta ^{(2r+1)}(0,z)$ majorizes for $\beta <\beta
_{2r+1}$ the density 
\begin{equation*}
\frac{\beta }{z}\rho ^{(r)}(\beta ,z)=\frac{\beta }{z}\rho _{\Lambda
}-\left( 2b_{\Lambda ,2}z+\cdots +2rb_{\Lambda ,2r}z^{2r-1}\right) ~,
\end{equation*}%
with the leading odd terms up to order $2r-1$ omitted. Because the r.h.s. of
equation (\ref{an}) depends on $a_{2j}(t)$, $j=1,\ldots ,r$, which diverges
as well as equation (\ref{a-2}) as $t_{0}\rightarrow \infty $, inequality (%
\ref{nbnCn}) for $n>2r$ do not hold without the modification we have made in
the previous section.

Let $F$ be given by (\ref{F}) with $n>2r$ and $\left\vert I\right\vert \leq
2r$ and define%
\begin{equation*}
a_{n}^{\emptyset }(t)=\sup_{x_{1}}\int \prod_{j=2}^{n}d^{2}x_{j}\frac{1}{%
2^{n}}\sum_{\mathbf{\sigma }\in \left\{ -1,1\right\} ^{n}}\left\vert \frac{%
\sigma _{1}+\cdots +\sigma _{n}}{n}\right\vert \left\vert \psi
_{n}^{c}(t,P_{n}\mathbf{\xi })\right\vert
\end{equation*}%
Note that $a_{n}^{\emptyset }\leq a_{n}$ and, by translational invariance,
the supreme over $x_{1}$ is irrelevant. As (\ref{mod}) holds with $%
a_{2}^{+} $ substituted by $a_{m}^{\emptyset }$, $m\leq 2r$, equation (\ref%
{an}) can be replaced by%
\begin{eqnarray}
a_{n}(t) &\leq &\frac{1}{2}\int_{0}^{t}ds\,e^{n\gamma (s,t)}\Gamma
(s)\left\{ \sum_{m=1}^{[n/2]}\binom{n}{m}\,m\,(n-m)\,a_{m}^{\emptyset
}(s)\,a_{n-m}(s)\right. \,~  \notag \\
&&+\left. \sum_{m=[n/2]+1}^{n-1}\binom{n}{m}\,m\,(n-m)\,a_{m}(s)\,a_{n-m}^{%
\emptyset }(s)\right\} ~  \label{an/0}
\end{eqnarray}%
where $[x]$ indicates the integer part of a real number $x$.

We now derive an integral equation for $a_{n}^{\emptyset }$'s similar to (%
\ref{an}). We assume that the improved stability condition (\ref{stb1})
holds. Applying the operation 
\begin{equation*}
\frac{1}{2^{n}}\sum_{\mathbf{\sigma }\in \left\{ -1,1\right\} ^{n}}\int
\prod_{i=1}^{n}dx_{i}\sum_{j=1}^{n}\sigma _{j}\dot{v}(t,\left\vert
x-x_{j}\right\vert )
\end{equation*}%
in both sides of (\ref{integral}) repeating the steps employed for (\ref{an}%
), with (\ref{B}) substituted by (\ref{stb1}), yields 
\begin{equation}
a_{n}^{\emptyset }(t)\leq \frac{1}{2}\int_{0}^{t}ds\,e^{(n-\delta
_{n})\gamma (s,t)}\Gamma (s)\sum_{m=1}^{n-1}\binom{n}{m}\,m\,(n-m)\,a_{m}^{%
\emptyset }(s)\,a_{n-m}^{\emptyset }(s)\,~  \label{an0}
\end{equation}%
with $\delta _{n}>1/n$ and $n=1,\ldots ,2r$. Here, as in equation (\ref{mod}%
), we have used translational invariance of $\dot{v}(t,\left\vert
x-y\right\vert )$ and, consequently, invariance of $\psi _{n}^{c}(\mathbf{%
\xi })$ under translation $\mathbf{\xi }\longrightarrow \mathbf{\xi }%
^{\prime }$ of all components $x_{j}^{\prime }=x_{j}+a$ by $a\in \mathbb{R}%
^{2}$.

By construction (see analogous equation (\ref{baA})), the Mayer
coefficients, regularized so that $b_{2j}=0$ if $j\leq r$, satisfy%
\begin{equation*}
nb_{n}\leq \left\{ 
\begin{array}{ccc}
a_{n}^{\emptyset }/(n-1)! & \mathrm{if} & n\leq 2r \\ 
a_{n}/(n-1)! & \mathrm{if} & n>2r%
\end{array}%
\right.
\end{equation*}%
and, to conclude the proof, we need to show%
\begin{equation*}
1+\sum_{j=2}^{2r}\frac{a_{j}^{\emptyset }(t)}{(j-1)!}z^{j-1}+\sum_{n=2r+1}^{%
\infty }\frac{a_{n}(t)}{(n-1)!}z^{n-1}\leq \Theta ^{(2r+1)}(t,z)
\end{equation*}%
which is equivalent to%
\begin{equation}
C_{n}\geq \left\{ 
\begin{array}{lll}
a_{n}^{\emptyset }/(n-1)!\equiv c_{n}^{\emptyset } & \mathrm{if} & 1<n\leq 2r
\\ 
a_{n}/(n-1)!\equiv c_{n} & \mathrm{if} & n>2r%
\end{array}%
\right.  \label{Ca}
\end{equation}%
where $\left( C_{n}\right) _{n\geq 2}$ , given by%
\begin{equation*}
\Theta ^{(k)}=1+\sum_{n=2}^{\infty }C_{n}z^{n-1}~,
\end{equation*}%
satisfy, in view of (\ref{ThetaRn}), the system of equations%
\begin{equation}
\dot{C}_{n}=\frac{k+1}{k}(n-1)\dot{B}C_{n}+\frac{n\Gamma }{2}%
\sum_{j=1}^{n-1}C_{j}~C_{n-j}  \label{CCn}
\end{equation}%
for $n=2,\ldots ,k=2r+1$ and 
\begin{equation}
\dot{C}_{n}=n\dot{B}C_{n}+\frac{n\Gamma }{2}\sum_{j=1}^{n-1}C_{j}~C_{n-j}
\label{Cn}
\end{equation}%
for $n>k$, with initial data $C_{n}(t_{0})=0$ if $n>1$ and $C_{1}(t)=1$ for
all $t\in \left[ t_{0},0\right] $.

Comparing (\ref{Theta}) with (\ref{ThetaRn}), we observe that the Lagrange
multiplier $L_{k}$, $k\geq 1$, leads to a modest improvement, from $n$ to $%
(k+1)(n-1)/k=n-(k-n+1)/k$, on the coefficient of the linear term $C_{n}$: 
\begin{equation}
n-1<n-\frac{k-n+1}{k}\leq n-\frac{1}{n}~.  \label{nnn}
\end{equation}

On the other hand, equations (\ref{an0}) and (\ref{an/0}) are equivalent to
the system%
\begin{equation*}
\dot{c}_{n}^{\emptyset }\leq (n-\delta _{n})\dot{B}c_{n}^{\emptyset }+\frac{%
n\Gamma }{2}\sum_{j=1}^{n-1}c_{j}^{\emptyset }~c_{n-j}^{\emptyset }
\end{equation*}%
for $n=2,\ldots ,2r+1$ and 
\begin{equation*}
\dot{c}_{n}\leq n\dot{B}c_{n}+\frac{n\Gamma }{2}\sum_{j=1}^{[n/2]}c_{j}^{%
\emptyset }~c_{n-j}+\frac{n\Gamma }{2}\sum_{j=[n/2]+1}^{n-1}c_{j}^{\emptyset
}~c_{n-j}
\end{equation*}%
for $n>r+1$ with the same initial data $c_{n}^{\emptyset }(0)=0$ if $1<n\leq
2r+1$, $c_{n}(0)=0$ if $n>2r+1$ and $a_{1}^{\emptyset }(t)=1$, from which,
together with (\ref{CCn}), (\ref{Cn}) and $\delta _{n}<1/n$, equations (\ref%
{Ca}) follow.

The Mayer series for the density of the Yukawa gas, with the singular part
of the norm of $\psi _{2j}^{c}$ for $j=2,\ldots ,r$ \ removed, converges if 
\begin{equation*}
\beta <\beta _{2r+1}\qquad \mathrm{and}\qquad \left\vert z\right\vert <\frac{%
\beta _{2r+1}-\beta }{\beta _{2r+1}\beta e}~
\end{equation*}%
with $\beta _{n}$ as in (\ref{betan}). This concludes the proof of Theorem %
\ref{singularity}.

\hfill $\Box $

We close this section with some comments on the improved stability condition.

\begin{remark}
\textit{\ \label{isc}}The non neutrality constraint is responsible for the
improvement (\ref{stb1}) since (\ref{B}) saturates if configurations with $%
\sigma _{1}+\cdots +\sigma _{n}=0$ are allowed. \textit{For }$n=2$, 
\begin{equation*}
\dot{U}_{2}(t;x_{1},x_{2},+,+)=\dot{U}_{2}(t;x_{1},x_{2},-,-)=\dot{v}\left(
t,|x_{1}-x_{2}|\right) \geq 0
\end{equation*}%
gives $\delta _{2}=2$ and this value exceeds $1/2$. For $n$ small, $\dot{U}%
_{n}(t;\mathbf{\xi })$ can be minimized over non neutral states $\mathbf{\xi 
}$ on a computer. Using the Mathematica software, the minimum for $n=3$%
\begin{eqnarray*}
\min_{\mathbf{\xi }}\dot{U}_{3}(t;\mathbf{\xi }) &\geq &2\min_{w\geq
0}\left( h(2w)-2h(w)\right) \dot{B} \\
&=&2\left( h(2w^{\ast })-2h(w^{\ast })\right) \dot{B}=-2.1162876...\cdot 
\dot{B}
\end{eqnarray*}%
is attained for a charge configuration disposed in a straight line with
alternate signal separated by an equal distance $w^{\ast }=0.4132466...$.
This gives a $\delta _{3}=0.8837124...$ larger than $1/3=0.3333333...$.

We have also found numerically the minimal configurations for non neutral
quadrupoles and quintupoles. For $n=4$ one of the exceeded charge is
expelled to infinity and the remaining charges dispose as in the $n=3$ case: 
\begin{equation*}
\min_{\mathbf{\xi }:\sum_{j}\sigma _{j}\neq 0}\dot{U}_{4}(t;\mathbf{\xi }%
)\geq 2\left( h(2w^{\ast })-2h(w^{\ast })\right) \dot{B}=-2.1162876...\cdot 
\dot{B}
\end{equation*}%
For $n=5$ the minimum is attained for a charge configuration disposed in a
straight line with alternate signal and symmetric with respect to the plane
intercepting the central charge perpendicular to the line:%
\begin{eqnarray*}
\min_{\mathbf{\xi }}\dot{U}_{5}(t;\mathbf{\xi }) &\geq &2\left(
h(2(w_{1}^{\ast }+w_{2}^{\ast }))+h(2w_{1}^{\ast })+2h(w_{1}^{\ast
}+w_{2}^{\ast })\right. \\
&&\qquad ~\qquad \left. -2h(2w_{1}^{\ast }+w_{2}^{\ast })-2h(w_{1}^{\ast
})-2h(w_{2}^{\ast })\right) \dot{B} \\
&=&-4.1879447...\cdot \dot{B}
\end{eqnarray*}%
where $w_{1}^{\ast }=0.4399588...$ and $w_{2}^{\ast }=0.2816336...$ are the
distances of the central charge to the neighbor and next neighbor charges,
respectively. In both case, we have $\delta _{4}=1.8837124...>0.25$ and $%
\delta _{5}=0.8120552...>0.2$.
\end{remark}

\begin{center}
{\large Acknowledgment }
\end{center}

\smallskip

We are grateful to Prof. Walter F. Wreszinki for instructive comments.


\begin{thebibliography}{PPNM}
\bibitem[Be]{Be} G. Benfatto. ``An iterated Mayer Expansion for the Yukawa
Gas'', J. Statist. Phys. \textbf{41}, 671--684 (1985)

\bibitem[BGN]{BGN} G. Benfatto, G. Gallavotti and F. Nicol\'o. ``On the
Massive Sine Gordon Equation in the First Few Regions of Collapse, Commun.
Math. Phys. \textbf{83}, 387 (1982)

\bibitem[B]{B} D. C. Brydges. ``A Short Course on Cluster Expansion'', in 
\textit{Critical Phenomena, Random Systems, Gauge Theories}, K. Ostervald and
R. Stora ed. (1986)

\bibitem[Br]{Br} D. C. Brydges. ``Functional Integrals and their
Applications'', notes for \textit{Troisi\`eme Cycle de la Physique en Suisse
Romande}, Lausanne (1992)

\bibitem[BF]{BF} David Brydges and Paul Federbush. ``A new Form of the Mayer
Expansion in Classical Statistical Mechanics'', J. Math. Phys. \textbf{19},
2064 - 2067 (1978)

\bibitem[BK]{BK} D. C. Brydges and T. Kennedy. ``Mayer Expansions and the
Hamilton-Jacobi Equation'', Journ. Stat. Phys. \textbf{48}, 19 - 49 (1987)

\bibitem[BK1]{BK1} D. C. Brydges and T. Kennedy. ``Convergence of Multiscale
Mayer Series'', in $VIII^{th}$ \textit{International Congress in
Mathematical Physics}, Marselle, World Scientific (1987)

\bibitem[BM]{BM} David C. Brydges and Ph. A. Martin. \textquotedblleft
Coulomb Systems at Low Density: A Review\textquotedblright , Journ. Stat.
Phys. \textbf{96}, 1163 - 1330 (1999)

\bibitem[FS]{FS} J\"{u}rg Fr\"{o}hlich and Thomas Spencer. \textquotedblleft
Phase Diagram and Critical Properties of (Classical) Coulomb
Systems\textquotedblright , Erice Lectures (1980)

\bibitem[GJ]{GJ} James Glimm and Arthur Jaffe. \textquotedblleft Quantum
Physics: A Functional Integral Point of View\textquotedblright ,
Springer-Verlag New York Inc. (1981)

\bibitem[GM]{GM} Markus G\"{o}pfert and Gerhard Mack. \textquotedblleft
Iterated Mayer Expansion for Classical Gases at Low
Temperatures\textquotedblright , Commun. Math. Phys. \textbf{81}, 97 - 126
(1981)

\bibitem[GN]{GN} G. Gallavotti and F. Nicol\'{o}. \textquotedblleft The
`Screening Phase Transitions' in the Two--Dimensional Coulomb
Gas\textquotedblright , Journ. Stat. Phys. \textbf{39}, 133 - 156 (1985)

\bibitem[I]{I} John Imbrie. \textquotedblleft Iterated Mayer Expansions and
their Application to Coulomb Gases\textquotedblright\ in \textit{Scaling and
self-similarity in physics}, Progr. Phys., \textbf{7}, 163--179. Birkh\"{a}%
user, Boston (1983)

\bibitem[K]{K} R. M. Corless, G. H. Gonnet, D. E. G. Hare, D. J. Jeffrey and
D. E. Knuth. \textquotedblleft On the Lambert W Function\textquotedblright ,
Adv. Comput. Math. \textbf{5}, 329-359 (1996)

\bibitem[LY]{LY} T. D. Lee and C. N. Yang. \textquotedblleft Statistical
Theory of Equation of States and Phase Transitions. II Lattice Gas and Ising
Model\textquotedblright , Phys. Rev. \textbf{87}, 404-409 (1952)


\bibitem[PPNM]{PPNM} Aldo Procacci, Emmanuel Pereira, Armando G. M. Neves
and Domingos H. U. Marchetti. \textquotedblleft Coulomb Interaction
Symmetries and the Mayer Series in the Two--Dimensional Dipole
Gas\textquotedblright , Journ. Stat. Phys. \textbf{87}, 877-889 (1997)

\bibitem[R]{R} David Ruelle. \textquotedblleft Statistical Mechanics:
Rigorous Results\textquotedblright , Addison-Wesley (1989)

\bibitem[S]{S} Barry Simon. \textquotedblleft The Statistical Mechanics of
Lattice Gases\textquotedblright , Princeton University Press (1993)

\bibitem[T]{T} L. Tonks. \textquotedblleft The Complete Equation of State of
One, Two and Three Dimensional Gases of Hard Elastic
Spheres\textquotedblright , Phys. Rev. \textbf{50}, 955 (1936)


\bibitem[YL]{YL} C. N. Yang and T. D. Lee. \textquotedblleft Statistical
Theory of Equation of States and Phase Transitions. I Theory of
Condensation\textquotedblright , Phys. Rev. \textbf{87}, 404-409 (1952)
\end{thebibliography}
\end{document}